\documentclass{revtex4}
\usepackage{amsmath} \usepackage{amsfonts} \usepackage{amssymb}
\usepackage{graphicx}
\usepackage{makeidx}         

\begin{document}

\def\jcmindex#1{\index{#1}}
\def\myidxeffect#1{{\bf\large #1}}

\title{sine-Gordon Equation:\\ From Discrete to Continuum}
\author{M. Chirilus-Bruckner}
\affiliation{%
School of Mathematics \& Statistics F07, University of Sydney, Sydney NSW 2006, Australia}

\author{C. Chong}
\affiliation{%
Department of of Mathematics and Statistics, University of
Massachusetts, Amherst, MA 01003-9305, USA}

\author{J. Cuevas-Maraver}
\affiliation{%
Nonlinear Physics Group, Departamento de F\'{\i}sica Aplicada I, Universidad de Sevilla. Escuela Polit{\'e}cnica Superior, C/ Virgen de \'Africa, 7, 41011-Sevilla, Spain
\\
Instituto de Matem\'aticas de la Universidad de Sevilla (IMUS).
Edificio Celestino Mutis. Avda. Reina Mercedes s/n, 41012-Sevilla, Spain}

\author{P.G. Kevrekidis}
\affiliation{%
Department of of Mathematics and Statistics, University of
Massachusetts, Amherst, MA 01003-9305, USA}

\begin{abstract}
In the present Chapter, we consider two prototypical Klein-Gordon models: the integrable sine-Gordon equation and the non-integrable
$\phi^4$ model. We focus, in particular, on two of their prototypical solutions, namely the kink-like heteroclinic connections and the
time-periodic, exponentially localized in space breather waveforms. Two limits of the discrete variants of these models are contrasted: on the one side, the analytically tractable original continuum limit, and on the opposite end, the highly discrete, so-called anti-continuum limit of vanishing coupling. Numerical computations are used to bridge these two limits, as regards the existence, stability and dynamical properties of the waves. Finally, a recent variant of this theme is presented in the form of $\mathcal{PT}$-symmetric Klein-Gordon field theories and a number of relevant results are touched upon.
\end{abstract}

\maketitle

\section{Introduction}

\jcmindex{\myidxeffect{D}!discrete and continuous models}
The sine-Gordon (sG) equation of the form \jcmindex{\myidxeffect{K}!Klein-Gordon PDEs}
\begin{eqnarray}
u_{tt}= u_{xx} - \sin(u) \, ,
\label{cheqn1}
\end{eqnarray}
with $ x,t,u=u(x,t) \in \mathbb{R}, $ is a prototypical mathematical model with a wide range of
applications~\cite{eilbeck,dauxois}; here, the subscripts $t,x$
denote partial derivatives with respect to time and space,
respectively.

One such example is the dynamics of
magnetic
flux propagation in Josephson junction (JJ)
transmission lines which are a promising way of
transmitting, storing and processing information.
The behavior
of such elements of magnetic flux (often referred to as fluxons)
can be accurately modeled by the propagation of kink-like,
heteroclinic solutions of the sine-Gordon equation. In fact,
even mechanical analogs based on systems of pendula
of such fluxon propagation
on a long Josephson junction, including adjustable torques
on the pendula (from air jets emulating the current of typical
JJ experiments) also exist~\cite{cirillo81}. In fact, the connection
of the sG equation and the JJ setting merits its own overview which
will be given in another Chapter of this Volume.

Another intriguing application of such a model
lies in the problem of charge density waves (CDWs)~\cite{gfdold4,gfdold5}.
Such CDWs originate from the fact that in several quasi-1d
conducting materials, it is favorable under a certain
temperature to undergo a phase transition to a state in which the
electron density develops small periodic distortions which
are followed by a modulation of the ion equilibrium positions.
As a result, a CDW condensate is formed which can be
subsequently pinned by impurities or interchain coupling.
In this setting too, the phase of the fluctuations of the
collective wavefunction of the condensate may encounter
a periodic potential and kink type solutions may thus be
suitable to model the behavior of a CDW condensate.

These are only some among the many examples where the
sine-Gordon continuum model has been used as a prototypical
system. On the other hand, a discrete variant of the model (DsG) \jcmindex{\myidxeffect{K}!Klein-Gordon lattices} \jcmindex{\myidxeffect{F}!Frenkel-Kontorova model}
\begin{eqnarray}
\ddot{u}_n=\epsilon \Delta_2(u_{n-1},u_n,u_{n+1}) - \sin(u_n)\, ,
\label{cheqn2}
\end{eqnarray}
with $ u_n = u_n(t) \in \mathbb{R}, n \in \mathbb{Z}, $ and where $ \Delta_2 $ is the central difference operator, is of particular interest in its own right; here
$\epsilon$ is the coupling between adjacent sites and
the subscript indexes the lattice sites. In that light,
the DsG has been originally proposed as a model for the dynamics
of dislocations in crystal-lattices, under the form of the
celebrated Frenkel-Kontorova model~\cite{FKpaper}; see also
the comprehensive book of~\cite{braun} and its discussion of
some of the historic origins of the model. The prototypical
realization of such a discrete system via an array of coupled
torsion pendula was originally proposed in the contributions
of~\cite{scott1}; see also~\cite{scott2}. This insightful connection has
provided a simple playground for a wide variety of research
studies that remains very active to this day exploring e.g.
the role of external driving and damping (including in the
stability of different types of breathing solutions of the
model)~\cite{uslars},  or that of longer-range interactions
and how they modify the nearest neighbor ones~\cite{hikihara}.

The sine-Gordon equation constitutes one of the integrable
nonlinear partial differential equations through the inverse
scattering transform~\cite{absegur}, which makes it a model
of particular interest in mathematical physics. 
However, it is also relevant to compare/contrast such a model
with variants that are {\it not} integrable, i.e., their
exact analytical solution cannot be prescribed on the basis
of suitable initial data in the general case. Perhaps one
of the most well-known Klein-Gordon examples of this
form is the so-called $\phi^4$ model~\cite{belova} of the form:
\jcmindex{\myidxeffect{K}!Klein-Gordon PDEs} \jcmindex{\myidxeffect{$\phi$}!$\phi^4$ model}
\begin{eqnarray}
u_{tt}=u_{xx} + 2 (u - u^3) \, .
\label{cheqn3}
\end{eqnarray}
This model has been physically argued as
being of relevance in describing domain walls in cosmological
settings~\cite{anninos}, but also structural phase transitions,
uniaxial ferroelectrics, or even simple polymeric chains; see,
e.g.,~\cite{campbell} and references therein. At this continuum
limit, one of the particularly intriguing features that were discovered
early on was the existence of a fractal
structure~\cite{anninos}
in the collisions between the fundamental nonlinear
waves, once again, a kink and an antikink in this model.
Notice here the fundamental contrast of such a fractal structure
with the completely inoccuous collisions arising in the sine-Gordon
integrable model, whereby the integrability and presence of an infinite
number of conservation laws dictates a completely elastic collision
outcome (and a mere phase-shifting of the kink or even breather
solitons).
This topic of the complex collisional outcomes
in the context of the $\phi^4$ model
was initiated by the numerical investigations of
Refs.~\cite{campbell,campbell2}
and was later studied in~\cite{belova,anninos} and is still under
active investigation see, e.g., the careful
recent mathematical analysis
of the relevant mechanism provided in Refs.~\cite{roy1,roy2}.

It should be added here that a discrete variant of the $\phi^4$ model
\begin{eqnarray}
\ddot{u}_n=\epsilon \Delta_2(u_{n-1},u_n,u_{n+1}) + 2 (u_n - u_n^3) \, ,
\label{cheqn4}
\end{eqnarray}
is also a model of both mathematical and physical interest. 
For instance, discrete
double well models arise in various physical settings such as
electronic excitations in conducting polymers~\cite{roy4}, structural
phase transitions in ferroic materials~\cite{roy5} i.e., on
crystal lattices in ferroelectrics, ferromagnets, and ferroelastics.\\

Since this Chapter considers a variety of equations, we summarize them in Table~1.

\begin{table}
\begin{center}
{\renewcommand{\arraystretch}{1.5}
\begin{tabular}{l|l}
(KG)  & \hspace{.2cm} $u_{tt}= u_{xx} - f(u) \, , \  x,t,u=u(x,t) \in \mathbb{R} $\\[.2cm]
\hline\\[-.2cm]
discrete (KG)   \hspace{.2cm}&\hspace{.2cm} $    \ddot{u}_n=\epsilon \Delta_2(u_{n-1},u_n,u_{n+1}) - f(u_n) \, , \  u_n=u(t) \in \mathbb{R}, n \in  \mathbb{Z}$\\[.2cm]
(AC) limit &\hspace{.2cm} $   \ddot{u}_n=  - f(u_n) $
\end{tabular}
}
\end{center}
\caption{The (KG) equation in its continuum, discrete and anti-continuum form. In this Chapter, we consider the sG variant where $f(u) = \sin(u)$
and the $\phi^4$ variant where $f(u) = -2(u-u^3)$.}
\end{table}

Our aim in the present Chapter is to present a view of the principal solutions -- kinks, breathers and solutions akin to them -- of two different Klein-Gordon (KG) equations: the one-dimensional sine-Gordon equation, $ f(u) = \sin(u) $, and its non-integrable counterpart, namely the $\phi^4$ model, $  f(u) = \alpha u + \beta u^3 $,  from the complementary perspectives of the continuum and the highly-discrete model, the so-called anti-continuum (AC) limit. As alluded to, our goal is to outline how the analytically more tractable continuum version and (AC) limit allows insight into the respective discrete version and how the different scenarios compare.  In that light, in section 2, we explore the continuum and discrete kinks, while in section 3, we focus on the continuum and discrete breathers, in both cases displaying a collection of numerical and analytical results concerning existence and (spectral) stability.  Finally, in section 4, we present a fairly timely alternative variant of the models in the form of $\mathcal{PT}$-symmetric Klein-Gordon field theories. These are models, which in the spirit of the original proposal of C. Bender and his collaborators for Schr{\"o}dinger models~\cite{bend1,bend2} (see also the review~\cite{bend3}) are invariant under the joint action of the symmetries of parity ($\mathcal{P}$) and time reversal ($\mathcal{T}$), potentially bearing in this way a real spectrum even for an ``open'' (i.e., bearing gain and loss) system.\\

\section{The Kink Case}

\subsection{The Continuum Limit and its Spectral Properties}

We start our considerations from the original continuum limit Eq.~(\ref{cheqn1}). This limit possesses standing
wave solutions that satisfy the ODE of the form
$u_{xx} - \sin(u)=0$. This ODE has homogeneous steady
states which are $u=0$ mod($2 \pi$) and unstable ones
which are $u=\pi$ mod($2 \pi$). The heteroclinic connections
that connect $0$ with $2 \pi$ are typically referred to as
kinks and their explicit functional form is given by
\jcmindex{\myidxeffect{K}!kink}
\begin{eqnarray}
u(x)=4 \arctan(e^{x-x_0}).
\label{cheqn5}
\end{eqnarray}
The presence of an undetermined constant $x_0$ reveals the
effective {\it translational invariance} of the model, 
according to which the kink can be centered equivalently at any point
$x_0$ along the one-dimensional line. In accordance with the
well established theory of Noether, this invariance (with respect
to translations) is tantamount to the existence
of a conservation law for the field theory (which in this case is the conservation of the linear
momentum $P=- \int u_t u_x dx$).  Another important conservation law among the infinitely many
of the sG equation is the conservation of energy (which corresponds to invariance under time shifts). In this context,
the energy or Hamiltonian
\begin{eqnarray}
H= \int \frac{1}{2} u_t^2 + \frac{1}{2} u_x^2 + \left(1 - \cos(u)\right) dx,
\label{cheqn6}
\end{eqnarray} is a constant of motion.
The D'Alembertian structure of the underlying linear
differential operator leads to invariance under the Lorentz transformations
of the form:
\begin{eqnarray}
x'=\gamma(x - v t), \quad \quad t'=\gamma(t- v x),
\label{cheqn7}
\end{eqnarray}
where $\gamma=1/\sqrt{1-v^2}$.
The equivalence of the wave operator in the original variables
$(x,t)$ vs. the new variables $(x',t')$ establishes that the
transformation of any of the above standing kinks centered at
$x_0$ enables the formation of a {\it traveling variant} of the
same solution
\begin{eqnarray}
u(x,t)=4 \arctan(e^{\gamma (x-x_0 - v t)}),
\label{cheqn8}
\end{eqnarray}
which propagates through the one dimensional line with speed $v$.
Given the equivalence of this traveling kink of the continuum
problem (upon the transformation~\eqref{cheqn7}) with the static kink, we will
focus predominantly on the latter in what follows.

One can linearize around static solutions of the sG to determine
the fate of small perturbations to such solutions. This can be
achieved using the ansatz $u(x,t)=u_0(x) + \epsilon w(x,t)$, whereby
$w$ satisfies the linear PDE
$w_{tt}= w_{xx} - \cos(u_0(x)) w$, which can be solved via separation
of variables $w(x,t)=e^{\lambda t} v(x)$, which leads to the eigenvalue
problem
\begin{eqnarray}
\lambda^2 v= \left[ \partial_x^2 - \cos(u_0) \right] v.
\label{cheqn9}
\end{eqnarray}
This linearized equation yields the above professed stability
of the $u_0=0$ and instability of the $u_0=\pi$ homogeneous
states, as the former leads to $\lambda= \pm i \sqrt{1+k^2}$
where $k$ is the wavenumber of a plane wave $v(x)=e^{i k x}$, while the latter leads to
$\lambda=\pm \sqrt{1-k^2}$. It is clear that the former
state will bear a continuous spectrum along the imaginary
axis of the spectral plane $(\lambda_r,\lambda_i)$ of the
eigenvalues $\lambda=\lambda_r + i \lambda_i$, while the
latter will feature a band of unstable eigenvalues arising
for wavenumbers $|k| < 1$.

\begin{figure}
\begin{tabular}{cc}
\includegraphics[width=6cm]{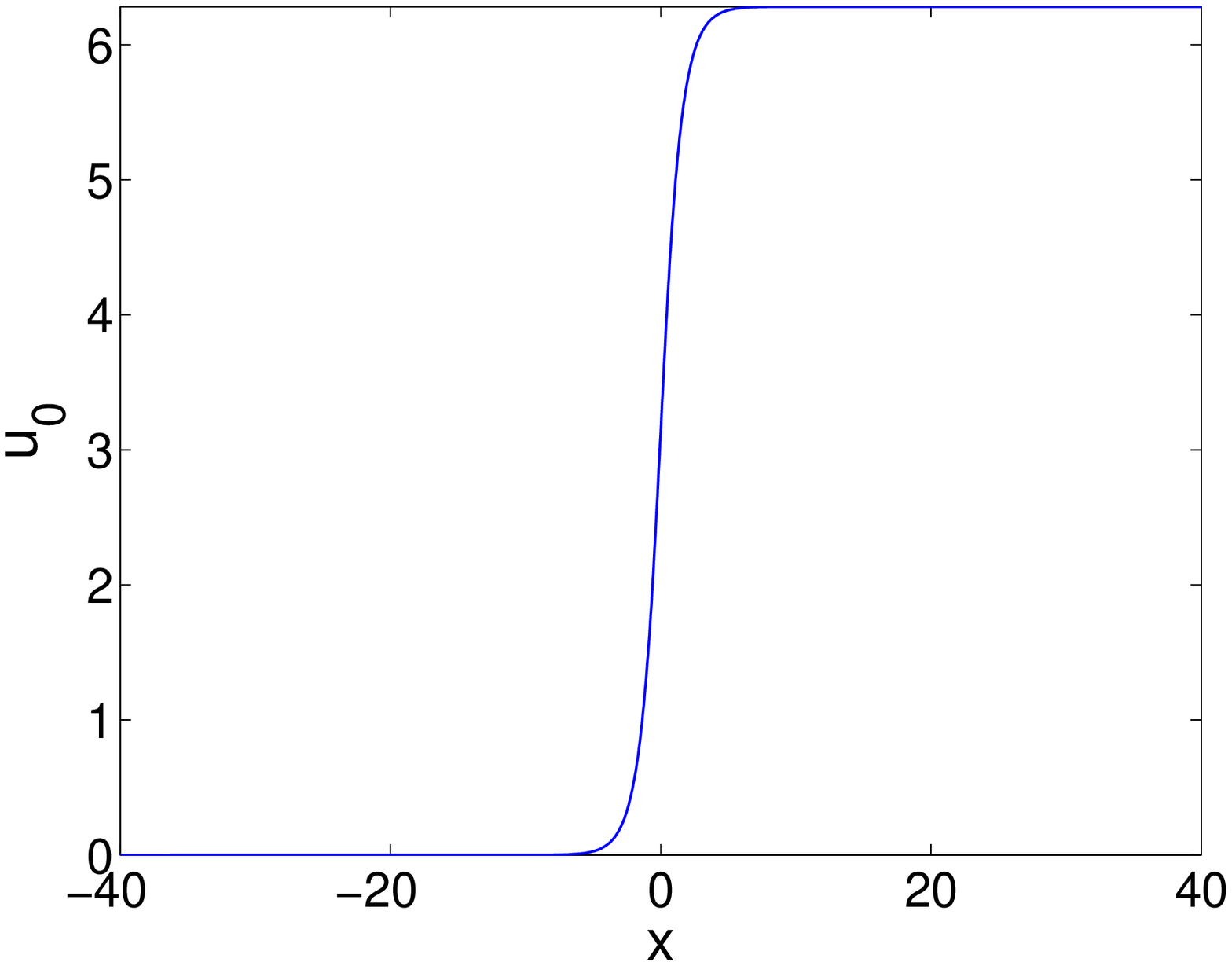}
\includegraphics[width=6cm]{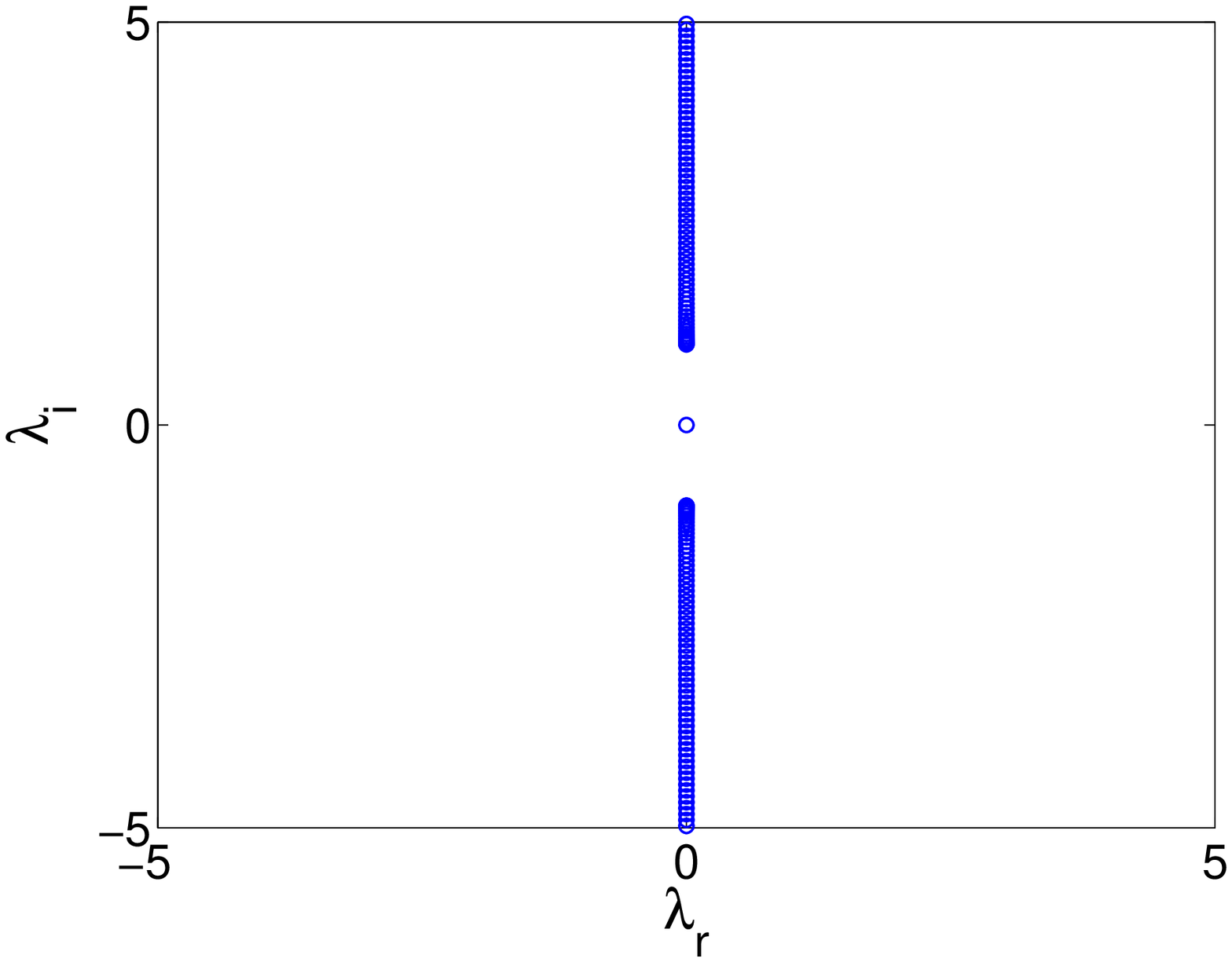}
\end{tabular}
\caption{The left panel shows the profile $u_0$ of the kink of the sG model
as a function of $x$.
The right panel shows the spectral plane $(\lambda_r,\lambda_i)$
of the linearization eigenvalues $\lambda=\lambda_r + i \lambda_i$; see
the detailed discussion in the text.}
\label{chfig1}
\end{figure}

Interestingly, the spectrum is explicitly available
even in the case of the kink, as it leads to a Schr{\"o}dinger
problem with a (special case of a)
Rosen-Morse type potential which is well-known from quantum mechanics.
The spectrum features a pair of eigenmodes
at $\lambda=0$, which are directly associated with the
translational invariance, since the corresponding eigenfunction
is $v=u_{0,x}=(1/\sqrt{2}) {\rm sech}(x-x_0)$ and arises directly from
the differentiation of the steady state equation $u_{0,xx}-\sin(u_0)=0$ (the derivative operator being the generator of the translation group). This mode is the only localized mode of the point spectrum
around the kink. The remainder of the spectrum consists of purely
continuous spectrum of non-decaying eigenfunctions whose explicit
form is given by:
\begin{eqnarray}
v^{(k)}=e^{i k (x-x_0)} \frac{k+i \tanh(x-x_0)}{\sqrt{2 \pi} (k+i)}.
\label{cheqn10}
\end{eqnarray}
The continuous spectrum of the linearization around the kink
shares exactly the same eigenvalues $\lambda= \pm i \sqrt{1+k^2}$
as the ones of the spectrum around $0$ or $2 \pi$ i.e., the
two homogeneous states that the kink heteroclinic orbit connects.
A typical example of the kink of the continuum sG equation and
of its corresponding spectrum is shown in Fig.~\ref{chfig1}.

In a sense, one may think of the spectrum of the kinks
as consisting of two separate ``ingredients''. On the one hand,
there is the point spectrum (which consists purely
of the neutral mode at $\lambda=0$) which pertains to the
coherent structure itself. On the other hand, there is the
continuous spectrum which corresponds to the background on which
the waveform ``lives''.  The background for the kink
is the $0$ state on the left and the $2 \pi$ on the
right.  For an anti-kink, where the opposite sign exists
in the exponent of the right hand side of Eq.~(\ref{cheqn5})
(and which connects $2\pi$ with $0$), this is reversed.
We note that the parsing of the spectrum into localized and extended
segments is of particular relevance when considering e.g.
the effect of collisions between kinks and anti-kinks and
how these reflect the integrability of the dynamics through
the well-known notion of elasticity of soliton collisions~\cite{pgk01}.

Results for kinks and their
spectrum in the case of the $\phi^4$ model are similar to the sG case. In particular,
in the case of Eq.~(\ref{cheqn3}), the kink assumes the
well-known form of the heteroclinic solution to the Duffing oscillator
\begin{eqnarray}
u_0(x)=\tanh(x-x_0).
\label{cheqn11}
\end{eqnarray}
Once again, the invariance with respect to translations,
and associated conservation law are present (a general feature
of Klein-Gordon field theories of the form $u_{tt}=u_{xx} - V'(u)$
for arbitrary field-dependent potentials $V(u)$). In this
case as well, the linearized problem can be written in
the form of an eigenvalue problem,
\begin{eqnarray}
\lambda^2 v=  \left[ \partial_x^2 + 2 (1-3 u_0^2) \right] v.
\label{cheqn12}
\end{eqnarray}
Here the homogeneous steady states are $\pm 1$ and $0$, with the
first two being stable with $\lambda = \pm i \sqrt{4 + k ^2}$,
while the latter is unstable with $\lambda = \pm \sqrt{2 - k^2}$.

\begin{figure}
\begin{tabular}{cc}
\includegraphics[width=6cm]{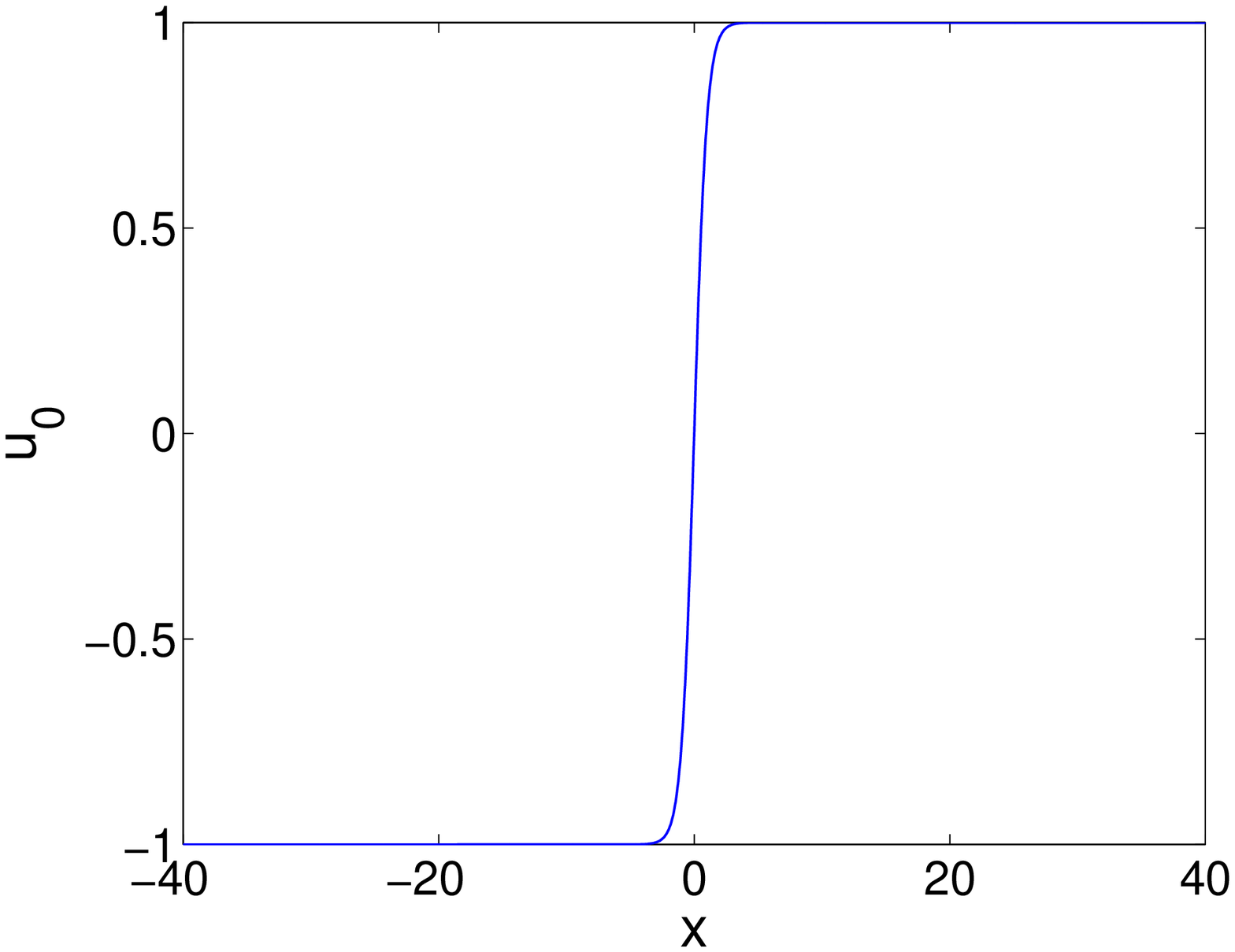}
\includegraphics[width=6cm]{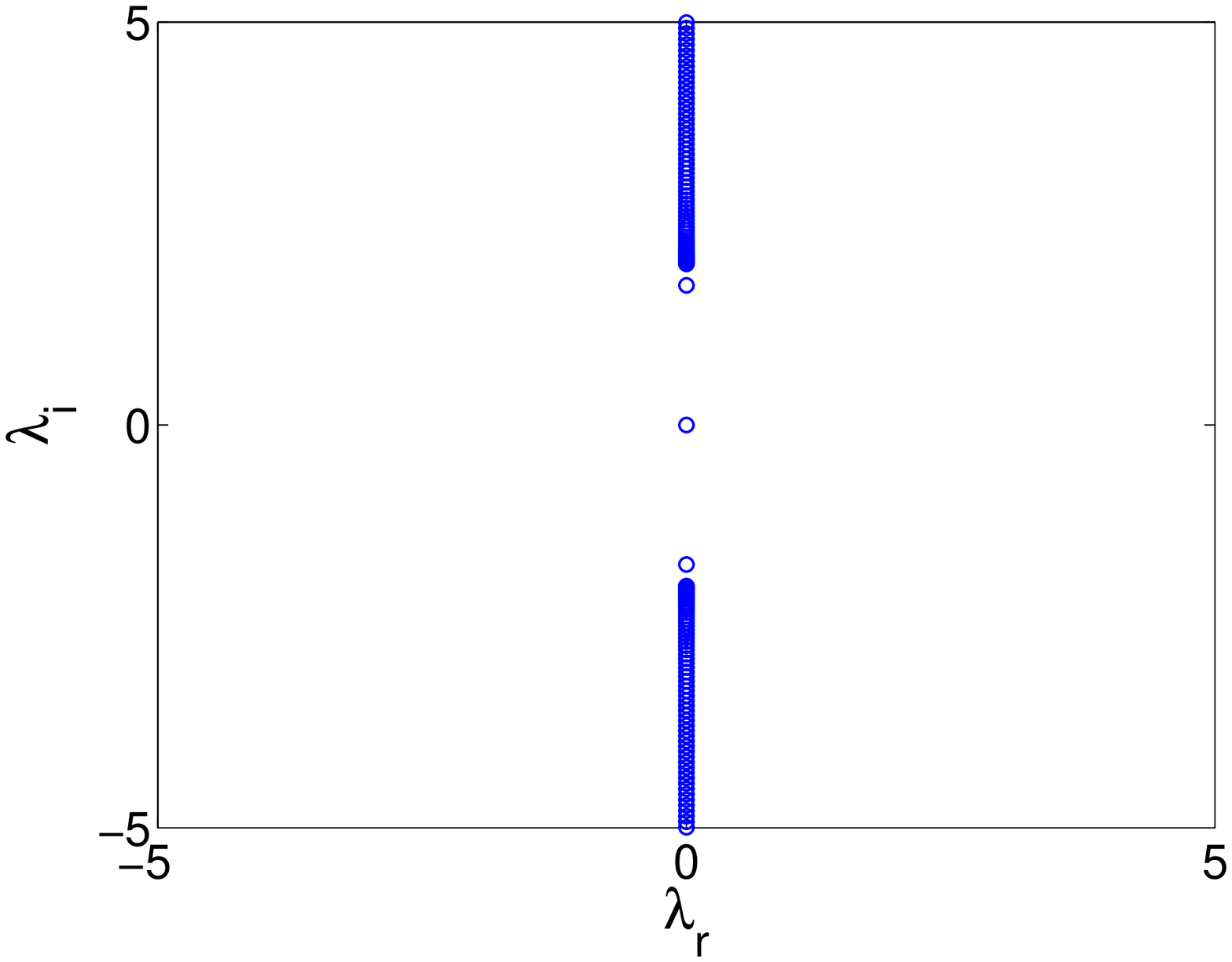}
\end{tabular}
\caption{Same as Fig.~\ref{chfig1}, but now for the kink
of the $\phi^4$ model. Notice on the right panel's spectral plane
the additional presence of the point spectrum eigenvalue
at $\lambda= \pm i \sqrt{3}$.}
\label{chfig2}
\end{figure}

The spectrum of the kink is also known in this case, although
it is more complex than that of the sG problem.
In particular, the neutral mode with $\lambda=0$ has an eigenfunction
$v(x)=\sqrt{3/4} {\rm sech}^2(x-x_0)$. The continuous spectrum
pertains to the band $\lambda = \pm i \sqrt{4 + k ^2}$ (again,
similarly to the underlying homogeneous problem) with the eigenfunctions
\begin{eqnarray}
v^{(k)}= N_k e^{i k x} \left(3 \tanh^2(x-x_0) -1-k^2 -3 i k \tanh(x-x_0)\right),
\label{cheqn13}
\end{eqnarray}
where $N_k^{-2}=4 \pi (2 (k^2/2+1)^2 + k^2/2)$~\cite{sugi}.
However, the key difference of the spectrum of the $\phi^4$ kink
from that of the sG is the presence of a point spectrum mode,
corresponding to a localized
eigenfunction, in the {\it gap} between the origin of the spectral plane
and the continuous spectrum (see Fig.~\ref{chfig2}). This internal (often referred to
also as shape) mode of the kink has an eigenvalue
$\lambda= \pm \sqrt{3} i$ and a corresponding eigenfunction,
\begin{eqnarray}
v(x)=\sqrt{\frac{3}{2}} \tanh(x-x_0) {\rm sech}(x-x_0).
\label{cheqn14}
\end{eqnarray}
It is this localized ``shape" mode that has been argued to be
typically responsible for the very rich phenomenology of collisions
of $\phi^4$ kinks and anti-kinks as explored e.g.
in~\cite{anninos,campbell,campbell2,roy1,roy2}, among other
works.

\subsection{Anti-Continuum Limit and its Spectral Properties}
\jcmindex{\myidxeffect{A}!anti-continuum limit}
We now turn to the opposite end of the continuum limit to construct
a complementary picture of the kink existence and stability properties,
so that we can connect the two in the next subsection.
In particular, we focus now on Eqs.~(\ref{cheqn2}) and~(\ref{cheqn4}),
which correspond to the discrete models. Here, one way to interpret
$\epsilon$ (e.g. in applications such as coupled torsion pendula)
is that of the spring constant linearly connecting adjacent such
pendula. A way of interpretation closer to numerical analysis is that
of adjacent nodes of a lattice, separated by a distance $\Delta x$,
such that $\epsilon=1/\Delta x^2$. This represents then a finite
difference scheme for the spatial discretization of the
corresponding continuum models of Eqs.~(\ref{cheqn1}) and~(\ref{cheqn3})
and as such the continuum limit is approached when $\Delta x \rightarrow 0
\Rightarrow \epsilon \rightarrow \infty$.

However, what was pioneered by MacKay and Aubry about 20 years
ago~\cite{macaub},
was a technique that offered a completely alternative viewpoint
to the one approaching the continuum as above. In particular, they
advocated using the {\it opposite} limit, namely that of
uncoupled oscillators at $\epsilon=0$. This was dubbed the anti-continuum (AC)
limit. There, our system consists of isolated oscillators of the
form $\ddot{u}_n=-\sin(u_n)$ for the sG and $\ddot{u}_n=2 (u_n-u_n^3)$
for $\phi^4$. The steady states now are {\it solely} $0$ and $\pi$
(mod$(2 \pi)$ as usual) and $\pm 1$ and $0$ for the two models, respectively.
Hence one can think of ways to ``initiate'' a kink at this limit,
which if $\epsilon$ starts becoming finite, upon continuation (using
a bifurcation theory analogy), the kink will start gradually picking
more sites progressively of more nontrivial ordinate. In this way, the
kink will be progressively ``fleshed out'' (i.e., the spine of the
heteroclinic connection will be populated) and will start looking
more like its continuum analog, which will be eventually traced
as $\epsilon \rightarrow \infty$. This kind of approach was
used e.g. in~\cite{balmf} to construct a variety of kink and
multi-kink solutions. Here, we will only explore the example
of states leading (as $\epsilon$ is increased) to the single kink,
and persisting from the AC to the continuum limit.

Upon some toying with the relevant background states, one realizes
that there are two prototypical possibilities for creating such a kink.
One of them concerns the so-called {\it intersite-centered} configuration,
initiated at the AC limit through a sequence
$(0,\dots,0,0,2 \pi, 2\pi,\dots, 2 \pi)$ in the sG and similarly
$(-1,\dots,-1,-1,1, 1,\dots, 1)$ in the $\phi^4$.
On the other hand, there is the {\it onsite-centered} configuration,
which incorporates a point of the unstable steady state in the middle,
namely $(0,\dots,0,0,\pi, 2 \pi, 2\pi,\dots, 2 \pi)$ in the sG and similarly
$(-1,\dots,-1,-1, 0, 1, 1,\dots, 1)$ in the $\phi^4$.

If we now examine such configurations in the
AC limit, it will be straightforward to infer their stability since the linearization
operator under the discretization becomes an infinite dimensional
matrix operator (or from a computational perspective it is restricted
to a finite but large number of nodes domain). In this setting
the linearization Jacobian
matrix will be tri-diagonal with the following diagonal and super/sub-diagonal
elements:
\begin{eqnarray}
J_{i,i}=-2 \epsilon - V'(u_i), \quad \quad J_{i,i+1}=J_{i,i-1}=\epsilon.
\label{cheqn15}
\end{eqnarray}
Here $V'(u_i)=-\cos((u_0)_i)$ for the sG and $V'(u_i)=2 (1-3 (u_0)_i^2)$
for the $\phi^4$. In the AC limit ($\epsilon=0$)  the matrix becomes
diagonal and it is evident that the eigenvalues $\lambda^2$ will be
negative if the kink involves only the stable steady states ($0,2\pi$
or $\pm 1$) and will have as many unstable eigenvalue pairs as many
sites there are at the unstable fixed points $\pi$ for the sG and
$0$ for the $\phi^4$.

Now, the key question becomes what can we expect in terms of the
kink stability as we depart from the AC limit, i.e., for finite $\epsilon$.
To determine this, we resort to a well-known theorem
for matrix eigenvalue problems, namely Gerschgorin's theorem~\cite{atkinson}.
In particular, if the order of the Jacobian matrix is $N$, the
eigenvalues $\lambda^2$ thereof belong to the sets
\begin{eqnarray}
Z_i=\{z \in C, \quad |z-J_{i,i}|<r_i, \quad {\rm with} \quad r_i=\sum_{j=1, j\neq i}^N |J_{i,j}| \}.
\label{cheq16}
\end{eqnarray}
Applying this to our setting, we find that $|\lambda^2 + 2 \epsilon + V'(u_i)|  \leq 2 \epsilon$, which in turn implies
\begin{eqnarray}
-V'(u_i) -4 \epsilon \leq \lambda^2 \leq  -V'(u_i).
\label{cheqn16}
\end{eqnarray}
In the case of the stable fixed points (at $0,2 \pi$ or at $\pm 1$,
respectively for the 2 models), this inequality properly outlines the
edges of the continuous spectrum. Hence, this indicates stability for $ N \rightarrow \infty $,
at least for small $\epsilon$ of the inter-site centered kink.

However a critical point is that it also establishes the instability of
the site-centered kink that {\it always} bears one site centered
at $\pi$ or $0$ in the sG or, respectively, the $\phi^4$ case.
For this eigenvalue, we have $1-4 \epsilon \leq \lambda^2 \leq 1$
for sG, while $2-4 \epsilon \leq \lambda^2 \leq 2$. Hence, this approach
{\it guarantees} the instability of the onsite-centered kink
for $\epsilon \leq 1/4$ for the sG problem and for $\epsilon \leq 1/2$
in the $\phi^4$ case.

We now turn to the merging of the two pictures in the subsection that
follows.

\subsection{Merging the two pictures}
\label{sec:kinkmerge}

\begin{figure}
\begin{tabular}{cc}
\includegraphics[width=6cm]{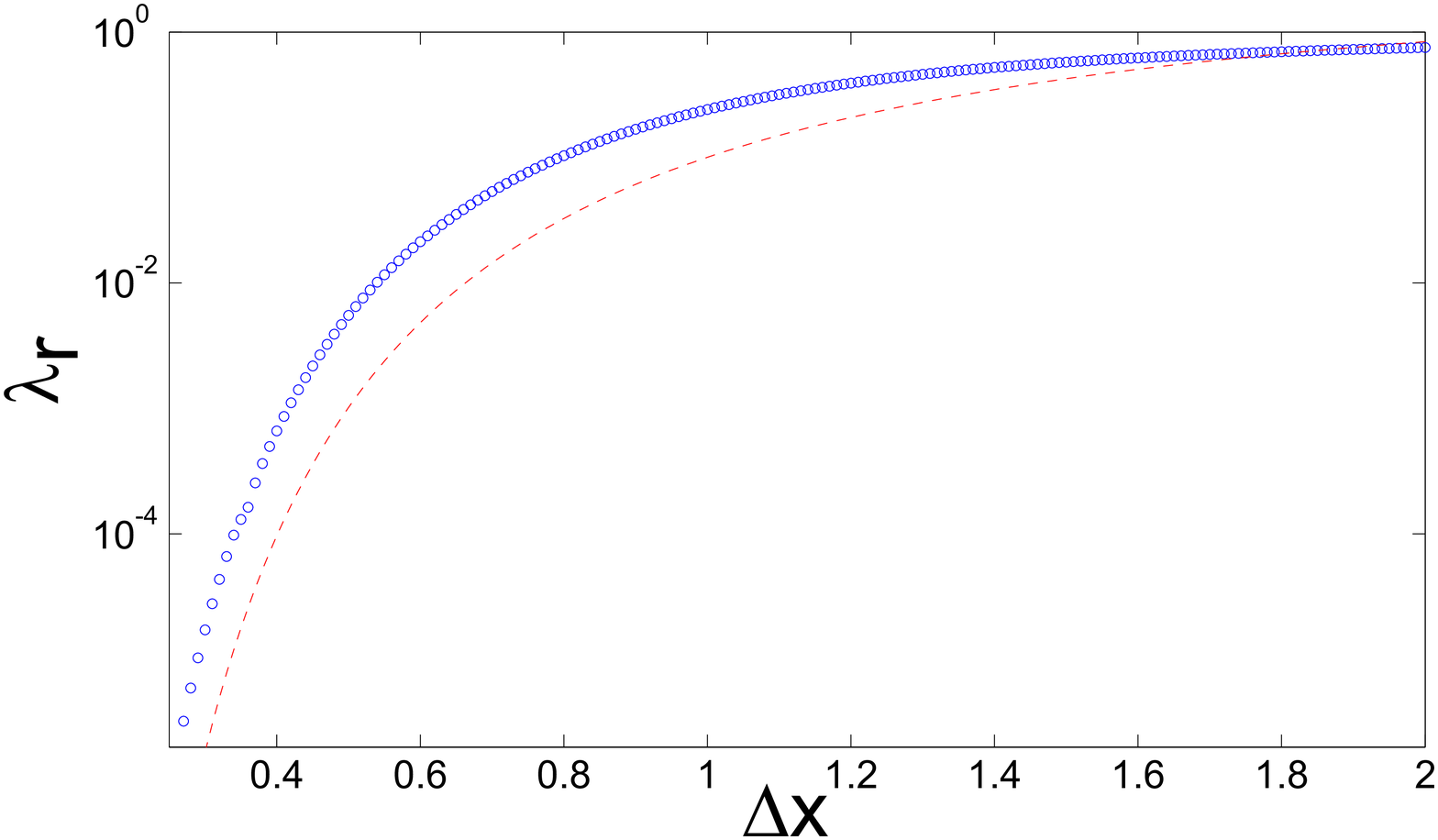}
\includegraphics[width=6cm]{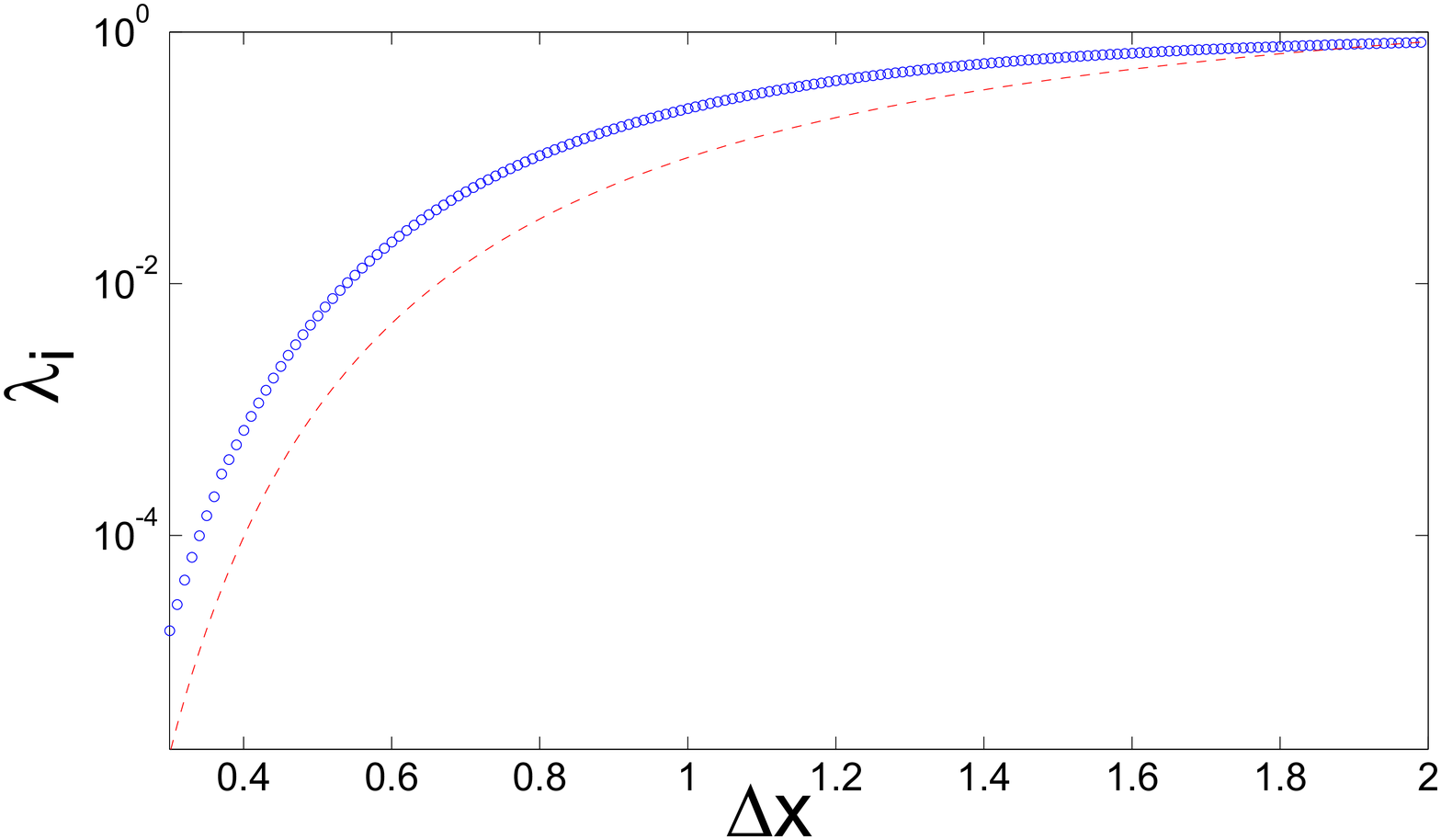}
\end{tabular}
\caption{The left panel shows the real eigenvalue pertaining
to the breaking of TI for the onsite centered kink. The blue
symbols denote the numerical computation of this eigenvalue and
the red dashed line provides the corresponding theoretical prediction
of Eq.~(\ref{cheqn19}). The right panel shows the imaginary eigenvalue
similarly associated to the breaking of TI for the stable
intersite centered kink (the symbols/lines have the same designation
as for the left panel).}
\label{chfig3}
\end{figure}

In the intermediate regime between the continuum limit of $\Delta x\rightarrow
0$ and $\epsilon \rightarrow \infty$ and the AC limit of
$\Delta x \rightarrow \infty$ and $\epsilon \rightarrow 0$, we
anticipate the following. For the site centered kink,
we expect that the unstable eigenvalue, while bounded away
from $0$ along the real line,
 according to our above estimate, will start approaching
the origin of the spectral plane as the continuum limit is approached.
While it is not guaranteed from our considerations herein that the eigenvalue will have a monotonic behavior,
we find that indeed it does, only arriving at the origin as
the asymptotic limit of the continuum regime is reached.
The approach of this eigenvalue towards the spectral plane
origin is shown in the left panel of Fig.~\ref{chfig3}.

On the other hand, in a similar manner, for the stable inter-site
centered kink, the approach towards the continuum limit needs to involve
an eigenvalue pair that approaches the origin as $\Delta x \rightarrow 0$
and the translational invariance is restored. Since in this
case all eigenvalues initially start at $\pm i$ (at $\epsilon \rightarrow 0$)
what occurs is that an eigenvalue pair bifurcates from the continuous
spectrum and starts approaching the origin of the spectral plane.
In this case too, although not necessarily so, the approach to the origin
is monotonic (and asymptotic) as $\Delta x$ is decreased.
This case is shown in the right panel of Fig.~\ref{chfig3}.

The eigenvalue associated to instability (resp. stability) of the site (resp. intersite) kinks
is related to the translational invariance (TI) in the continuum limit. It is the breaking
of the translational invariance once the system becomes discrete that is ``responsible"
for the instability/stability of these two different kinks.  The difference in the energies associated to these two kinks is defined as the celebrated Peierls-Nabarro (PN) barrier,
which is discussed at considerable length in the physical
literature~\cite{braun} as the barrier needed to be overcome for
a dislocation to move by a lattice site.

One of the typical
issues that arise in the calculation of the PN barrier is that it is
not straightforward to perturb off of the continuum limit in
the direction of discreteness. This problem was bypassed in~\cite{kk01}
by perturbing off of an exceptional (translationally invariant)
discretization of the sG model originally proposed by Speight
and Ward in~\cite{sw94}. This was in the form:
\begin{eqnarray}
\ddot{u}_n &=& 2 \epsilon \left( \sin(\frac{u_{n+1}-u_n}{2})
- \sin(\frac{u_n - u_{n-1}}{2}) \right) \\
\nonumber
 &-& \frac{1}{2} \left( \sin(\frac{u_{n+1} + u_n}{2})
+ \sin(\frac{u_n + u_{n-1}}{2}) \right).
\label{cheqn18}
\end{eqnarray}
The fact that this model is TI implies that it possesses kinks
which can be located arbitrarily, as is in fact evidenced by
the exact solutions $u_n=4 \arctan(e^{a (n-\xi)})$
where $a=\log((2+\Delta x)/(2 - \Delta x))$ and $\xi$ is a
free parameter. It also suggests that the relevant TI eigenvalue
pair is at $\lambda^2=0$, yet because of its inherent discrete
nature, the model is amenable to being perturbed to be reshaped
in the form of Eq.~(\ref{cheqn3}). Following this path,~\cite{kk01}
predicted that the relevant eigenvalue should be given by
\begin{eqnarray}
\lambda_r \approx \pm \frac{13.96}{\Delta x^{1/2}} e^{-\frac{\pi^2}{\Delta x}},
\quad \quad \lambda_i \approx \pm i \frac{13.96}{\Delta x^{1/2}} e^{-\frac{\pi^2}{\Delta x}}.
\label{cheqn19}
\end{eqnarray}
The first of these expressions applies to the real eigenvalue of
the onsite-centered kink, while the second to the imaginary one
of the intersite-centered kink. Both are applicable when approaching
the continuum limit. Yet, it was recognized in~\cite{kk01} that while
the functional form of the dependence should be suitable, the prefactor
may sustain additional contributions from higher order terms which
should appear in the same order of eigenvalue dependence. It is for
this reason that the dashed lines describing this theoretical dependence
of Eq.~(\ref{cheqn19}) capture the appropriate functional dependence
as the limit is approached, but clearly the prefactor is smaller
than it should be.

Entirely similar features also arise for
the $\phi^4$ case, which is thus not examined further here.
Instead, we now turn to the corresponding examination of
the breather waveform.

\section{The Breather Case}

\jcmindex{\myidxeffect{B}!breather}
The other fundamental solution of the sG model is a state
which is also exponentially localized in space (similarly
to the kink) but periodically varying in time. This is the
so-called breather state whose exact profile is given by
\begin{eqnarray}\label{EQ:breather_formula}
u(x,t)=4 \arctan \left(\sqrt{\frac{1-\omega^2}{\omega}} \sin(\omega (t-t_0))
{\rm sech}(\sqrt{1-\omega^2} (x-x_0) ) \right).
\label{cheqn20}
\end{eqnarray}
Aside from translations with respect to space and time
(discussed in the previous section), this solution
has a free parameter $\omega$, associated with the frequency of its
``breathing'' satisfying $0 < \omega < 1$. Interestingly,
the breather seems like the result of the merger of a kink and an
antikink, however although the energy of the kink and antikink is $8$ such
that the sum of their energies is $E=16$, the energy of
their bound state, i.e. the breather, is always less than that i.e.,
$E=16 \sqrt{1-\omega^2}$.

\subsection{Continuum limit: Genuine breathers and modulating pulses}
\jcmindex{\myidxeffect{B}!B\"acklund Transformation}
For the sG equation, which is an infinite dimensional completely integrable Hamiltonian system, breather solutions can be obtained through its auto-B\"acklund transformation (BT). This procedure is, of course, limited to integrable systems and does not help in the quest for breathers in general (KG) equations
\begin{align}\label{EQ:KG}
u_{tt} = u_{xx} - u + f(u) \, ,
\end{align}
with $ x,t,u=u(x,t) \in \mathbb{R}, $ and some nonlinearity $ f(u) $. To this end, various techniques have been employed and developed during the past decades to elucidate the existence of breathers or solutions akin to it. Roughly speaking, one can distinguish between three approaches. The first approach is the analysis via series expansion techniques (see \cite{kruskalseg, kichenassamy}) and the study of perturbed sG equations (see \cite{birnir_mackean_weinstein, denzler_93, denzler_97}), whose findings indicated the non-persistence of breather families and the continuation of breathers as solutions which lack the spatial localization property. Out of these results emerged the more general second approach of constructing  for \eqref{EQ:KG} so-called {\it modulating pulses}, which are moving breathers that are not necessarily spatially localized, but feature oscillating tails; see e.g. \cite{groves_schneider, shatah_zeng}. Yet a different, third approach exploits the wave packet structure of breathers. In
their small-amplitude
limit they can be thought of as wave
packets whose envelope is described by a Nonlinear Schr\"odinger (NLS) equation (see \cite{KirrmannSchneiderMielke} for a rigorous approximation
result). While perturbation
arguments in the first approach rely heavily on integrable systems techniques and/or the explicit solution formula known for breathers,
the second approach follows the realm of infinite dimensional dynamical systems and inspires an extension of center manifold theory beyond the classical case. The methodology of the third approach belongs to the theory of reduction to amplitude/modulation equations and plays an important role far beyond the setting discussed here.

\begin{figure}
\begin{center}
\begin{minipage}{4cm}
\begin{align*}
\tan\left( \frac{u + U}{4} \right) = \left( \frac{ a_1 + a_2 }{ a_1 - a_2 } \right) \, \tan\left( \frac{u_1 + u_2}{4} \right)
\end{align*}
\end{minipage}
\hspace{1cm}
\begin{minipage}{.0cm}
    \scalebox{.3}{\includegraphics{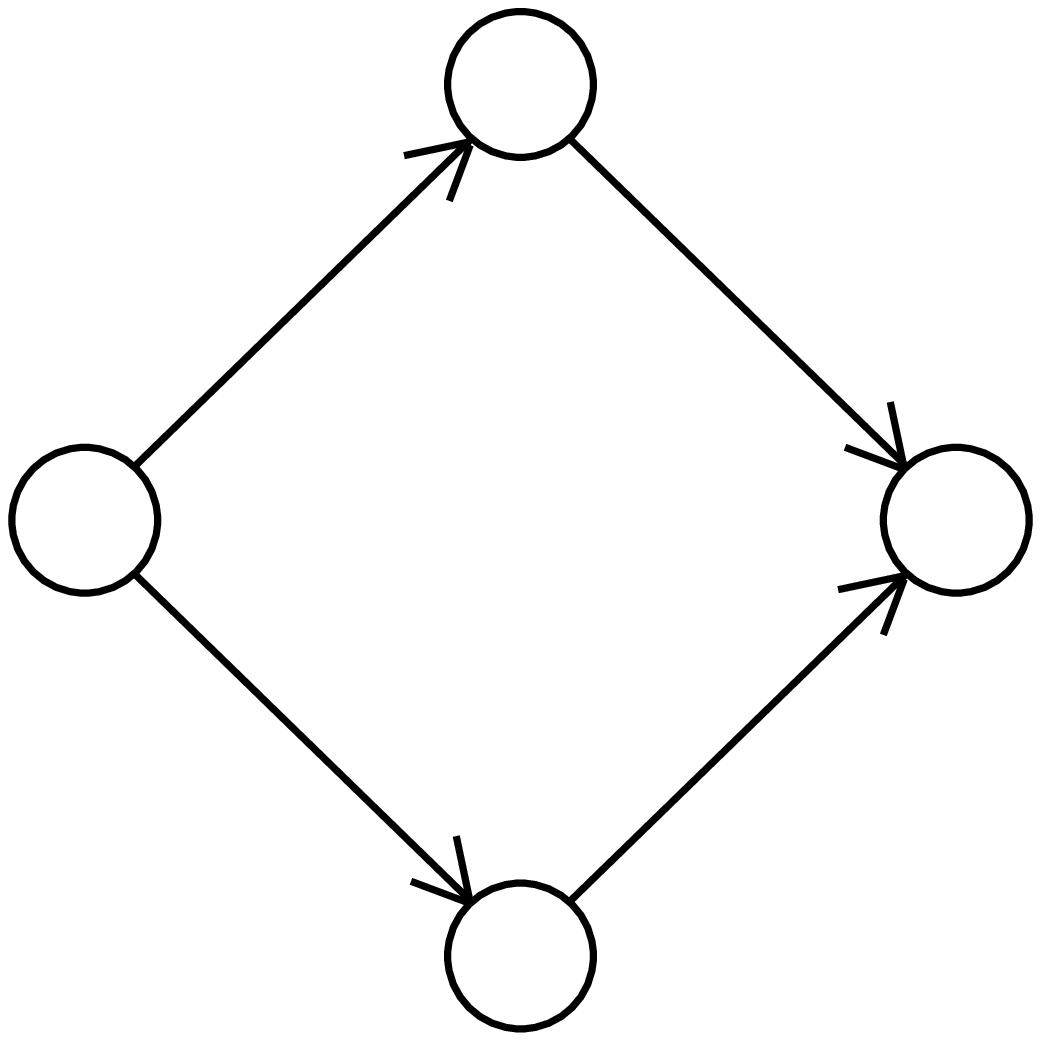}}\\
    \begin{picture}(0,0)
        \put(16,80){$ a_1 $}
        \put(71,80){$ a_2 $}
        \put(14,30){$ a_2 $}
        \put(71,30){$ a_1 $}
        \put(6,54){$ u $}
        \put(81,54){$ U $}
        \put(42,92){$ u_1 $}
        \put(42,17){$ u_2 $}
    \end{picture}
\end{minipage}
\caption{Scheme of commutativity of the BT: $ B( B ( u , a_1 ) , a_2  ) = B ( B ( u , a_2 ) , a_1 ) $}
\label{FIG:BT}
\end{center}
\end{figure}

Before we review each approach in more detail, let us briefly summarize the sG-breather construction. As alluded to, the sG equation admits an auto-B\"acklund transformation, which is, loosely speaking, a means of obtaining solutions to the sG equation by a procedure that is in some sense easier than solving a PDE (i.e. it might involve solving ODEs or algebraic equations). To be more precise, the BT maps ``solutions to solutions'', i.e., it is a mapping $ B $ with
\begin{align*}
B(u_1; a ) = u_2 \, , \qquad  u_1, u_2 \ \mathrm{solutions} \ \mathrm{of} \ \mathrm{(sG)}
\end{align*}
where $ a $ is some parameter. One can, for instance, use the trivial solution as a seed to grow non-trivial solutions like kinks or anti-kinks
\begin{align}\label{EQ:BT_kinks}
B(0,1) = 4 \, \arctan (e^x) \, , \qquad B(0,-1) = 4 \, \arctan (e^{-x}) \, .
\end{align}
It is even possible to derive a surprisingly simple relation between four different solutions  $ u,U,u_1,u_2 $ of the (sG) equation by using the commutativity theorem schematically depicted in Figure~\ref{FIG:BT} (cf., for instance, \cite{mclaughlin_scott}).

\begin{figure}[!ht]
 \centering
 \scalebox{.4}{\includegraphics{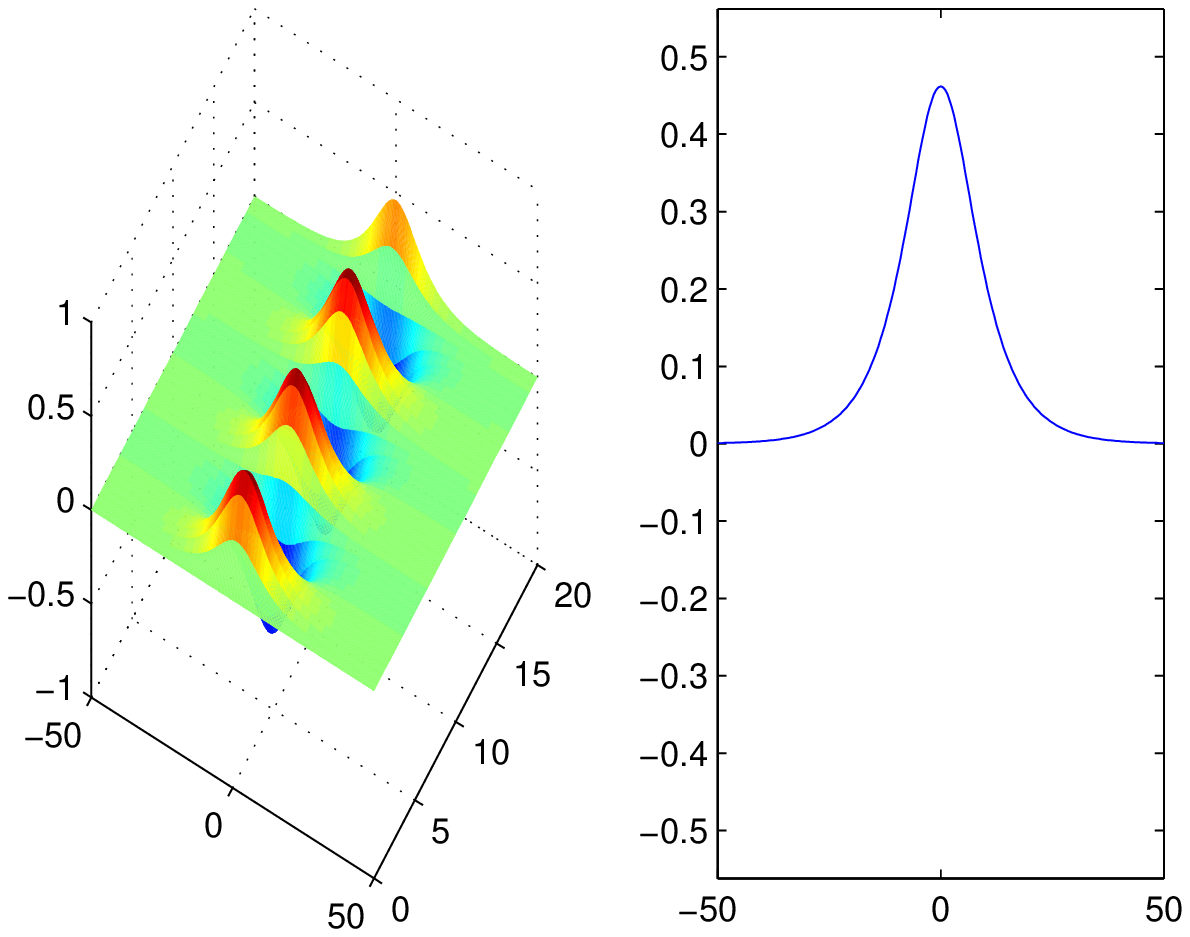}}
 \scalebox{.4}{\includegraphics{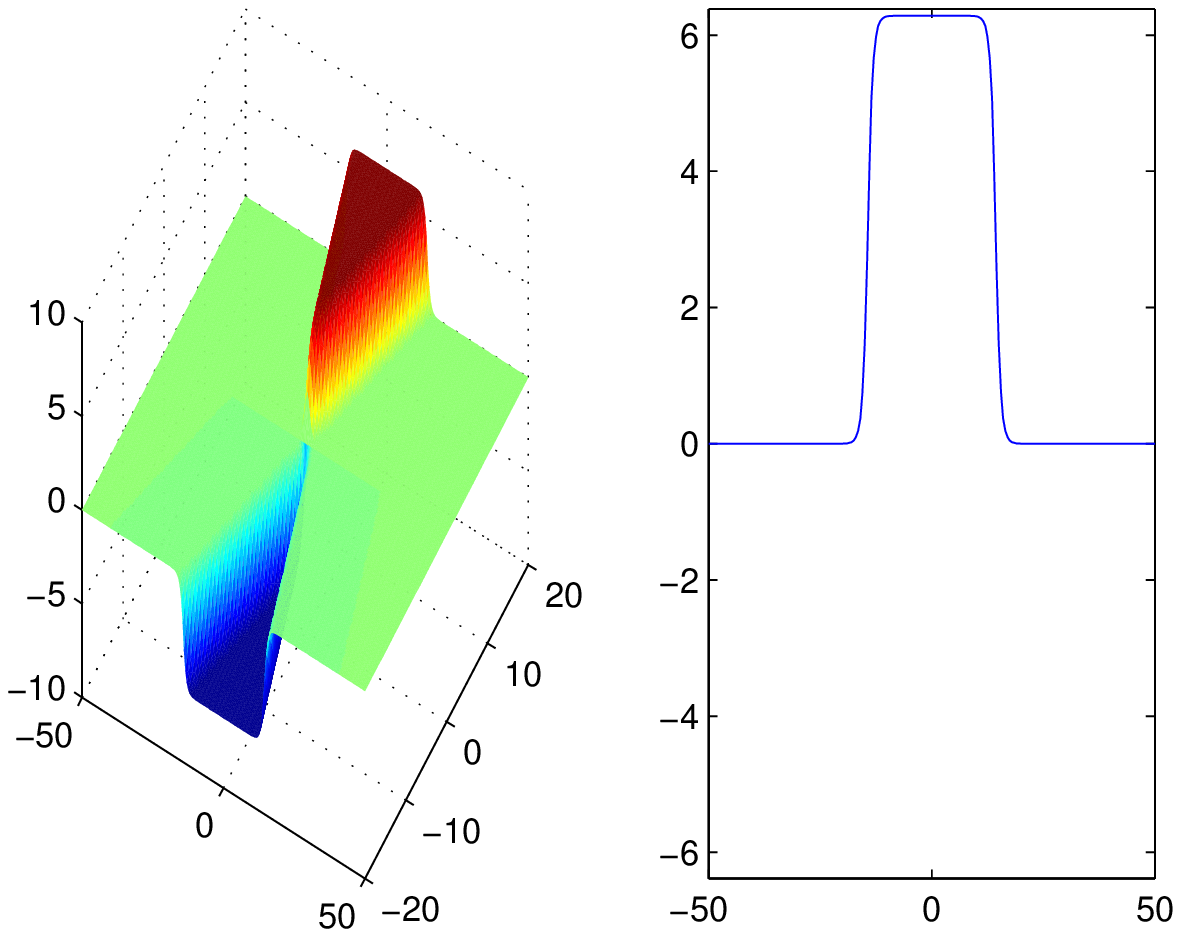}}
 \caption{{\it Top panel}: Standing breather with $ \omega = 0.99 $ in \eqref{EQ:BT_breather}. {\it Bottom panel}: Kink/anti-kink interaction with $ \omega = i 0.99 $ in \eqref{EQ:BT_kink_anti_kink_interaction}.}
  \label{FIG:BT_breather_kink_anti_kink}
\end{figure}

Using this scheme, one can derive a formula for (a family of) standing breathers (see Figure~\ref{FIG:BT_breather_kink_anti_kink})
\begin{align}\label{EQ:BT_breather}
 u_*(x,t;\omega) = 4 \, \arctan \left( \frac{m}{\omega} \, \frac{\sin(\omega t) }{ \cosh(m x) }  \right) \, , \quad m^2 + \omega^2 = 1 \, ,
\end{align}
with $ u = 0, u_{1/2} = \mathrm{arctan} \left( e^{ mx \pm i \omega t  } \right) $ and $ U = u_* $. Note that choosing $ \omega = i \widetilde{ \omega } \in i \mathbb{R} $ turns $ u_* $ into
\begin{align}\label{EQ:BT_kink_anti_kink_interaction}
u_*(x,t;i \widetilde{\omega} ) = 4 \, \arctan \left( \frac{m}{i \widetilde{ \omega } } \, \frac{\sin( i \widetilde{ \omega } t) }{ \cosh(m x) }  \right)= 4 \, \arctan \left( \frac{m}{ \widetilde{ \omega } } \, \frac{\sinh( \widetilde{ \omega } t) }{ \cosh(m x) }  \right),
\end{align}
which describes a kink/anti-kink solution (see Figure~\ref{FIG:BT_breather_kink_anti_kink}). Iterating this procedure results in more complicated solutions given by complexes of kinks and/or breathers.

As alluded to, one can derive kink solutions by a completely different viewpoint, namely, by making a traveling wave ansatz
$  u(x,t) = y\left(x - ct\right) \, , 0 \leq c < 1 \, ,$ and examining the traveling wave ODE whose heteroclinics are given by $  y(\xi) = 4 \, \arctan( e^{\pm \xi} ) \, ,\xi =  \gamma (x-ct) \, , \gamma = ( 1-c^2 )^{-1/2} $. For $ c = 0 $ we recover \eqref{EQ:BT_kinks}, while
the traveling kink can be viewed also as a consequence of
the Lorentz transformation \eqref{cheqn7}, as discussed above. Note, however, that breathers (or, more generally, any multikink solution) cannot be obtained from the traveling wave ODE since they are not stationary in any fixed co-moving frame. Hence, breathers can be viewed as genuinely infinite dimensional structures. While the explicit expressions \eqref{EQ:BT_breather} or \eqref{EQ:BT_kink_anti_kink_interaction} rely heavily on PDE tools for integrable systems, the general non-integrable case calls for different PDE techniques. Breather existence could, for instance, be explored via the construction of small-amplitude wave packets, which can be approximately performed for a large class of Klein-Gordon equations \eqref{EQ:KG} via reduction to amplitude/modulation equations. Despite their widespread use, rigorous justifications of such procedures are more sparse, with an early example being \cite{KirrmannSchneiderMielke}, where it was demonstrated that there are solutions $ u $ to the (sG) equation for
which
$$ \sup_{t \in \left[0, \frac{T_0}{\varepsilon^2} \right] } \, || u(\cdot,t) - u_{\rm a}(\cdot,t)  ||_{C_b^r} \leq C \varepsilon^{3/2},  $$
with $ 0 < \varepsilon \ll 1 $ being a small perturbation parameter and where $ u_{\rm a} $ is a small-amplitude wave packet
\begin{align}\label{EQ:wavepacket_ansatz}
 u_{a}(x,t) = \varepsilon \, A( \varepsilon (x-\omega'(k_0)t), \varepsilon^2 t )\, e^{ i k_0 x - i \omega(k_0) t} + c.c. + h.o.t. \, ,
\end{align}
with $ \omega(k) = \sqrt{k^2 +1}  $; the envelope $ A  = A(X,T) \in \mathbb{C} $ is obtained by solving the Nonlinear Schr\"odinger (NLS) equation
\begin{align}\label{EQ:NLS}
 2i \omega(k_0) \partial_T A = (\omega'(k_0)^2 - 1) \partial^2_X A - 3 |A|^2 A \, .
\end{align}
This result is also valid for general KG equations of the form given in~\eqref{EQ:KG} since small-amplitude solutions do not feel to leading order the exact structure of the nonlinearity. Note that, if we choose the NLS soliton as an envelope, the wave packet ansatz $ u_{a} $ resembles the traveling sG breather in the small-amplitude limit $ 0 < m = \varepsilon \ll 1 $. Note that, although the approximation result treats a large class of solutions  -- namely, those whose envelopes are given by solutions of the NLS --  of which the breather is just a special case, the limited time-validity of the construction (namely, $ t \in \left[0, \frac{T_0}{\varepsilon^2} \right] $) does not give the optimal result (that is, global existence in time) in the breather case. Such an improvement of the perturbation procedure has so far not been reported in the literature and would
be a particularly interesting direction for future work.

The analogy between breathers and general wave packets becomes even more remarkable when one considers interaction. Using the BT, one can construct, for instance, a 2-breather describing the elastic collision of two individual breathers. Such an interaction for small amplitude
wave packets in the general KG setting \eqref{EQ:KG} has striking similarities to the breather interaction: to leading order, the shape of the wave packets remains unchanged after collision, the only interaction effects being a shift of envelope and carrier waves (see \cite{ccsu} for KG with constant coefficients and \cite{cs} for KG with spatially periodic coefficients). In the case of large
amplitudes, however, the situation is far more complex, as is attested also by the complexity of interactions of simpler structures such as kinks in non-integrable KG models~\cite{belova,anninos,campbell,campbell2,roy1,roy2}. Since a large class of KG equations support (approximate) small-amplitude wave packets which seem akin to the sG breather, it had been long speculated that
exact breathers may exist for other KG equations as well. This belief has
been proven wrong in numerous case examples
from several different viewpoints.\\

The work of \cite{kruskalseg} contrasted the
sG setting with the $ \phi^4 $ model.
By employing the Fourier representation
\begin{align}\label{EQ:fourier_representation}
 u(x,t) = \sum_{n \in \mathbb{Z}} \, \hat{u}_n(x) \, e^{ in \omega_*  t },
\end{align}
for time-periodic solutions with some temporal frequency $ \omega_*/2 \pi $, they derive an infinite-dimensional system of equations for the Fourier coefficients $ \hat{u}_n $ and conclude that only the sG equation allows spatially localized solutions $ \hat{u}_n(x) \rightarrow 0, x \rightarrow \pm \infty $, which in turn give rise to breathers, while the $ \phi^4 $ model only supports approximate breathers that slowly radiate their energy to $ \pm \infty $.

The work of \cite{kichenassamy} considered a more general KG setting with
a force term $f(u)$ in \eqref{EQ:KG} and presented a detailed convergence analysis of different breather series
expansions relating the corresponding expansion coefficients to the coefficients $ f_n $ in $ f(u) = \sum_{n = 1}^{\infty} f_n \, u^n$. The main result singles out the cases $ f(u) = Cu, C \sinh(au), C \sin(au) $ which are the only choices yielding analyticity of the solution, while (again) only $ f(u) = C \sin(au) $ gives breathers. In the early 90's,
the works of \cite{birnir_mackean_weinstein} and
\cite{denzler_93,denzler_97} studied perturbed sG equations
\begin{align}\label{EQ:perturbed_sG}
  u_{tt} - u_{xx} + \sin(u) = \varepsilon P(u;\varepsilon) \, , \quad 0 < \varepsilon \ll 1 \, ,
\end{align}
and concluded that breathers can persist (in families) if and only if the perturbation is equivalent to a rescaling of the sG equation. Consequently, the non-persistence results indicated that breathers perturb into solutions that are similar in nature, but not localized. This intuition was confirmed in~\cite{groves_schneider}
(and later also in~\cite{shatah_zeng}) by using spatial dynamics and a blend of invariant manifold, partial normal form and bifurcation theory for KG
equations \eqref{EQ:KG}. They study so-called modulating pulse solutions,
that is,
\begin{align*}
 u(x,t) = v ( x- (1/c_{\rm ph}) t; x- \left(c_{\rm ph} + \gamma_1 \varepsilon^2 \right) t ) = v(\xi, y),
\end{align*}
where $ v $ is periodic in $ y $ with period $ 2\pi/k_0 $ for some $ k_0 > 0 $. Note that the sG-breathers fall into this class with the additional property of being localized in $ \xi $, a feature that seems rather exceptional in the following way. The infinite dimensional system for $ (v,v_x) $ on the phase space  $ \chi = \{ (v,w) \in H_{\rm per}^{s+1} \times H_{\rm per}^{s}  \}, s \geq 0, $ (where now $ x $ is the dynamic variable, hence, the name spatial dynamics and $ H_{\rm per}^{s} $ are $ L^2 $-based Sobolev spaces of periodic functions) has a Hamiltonian structure whose linearization features infinitely many eigenvalues on the imaginary axis and exactly two real eigenvalues $  \lambda_{1,2} =  \pm \varepsilon \rho $.
The leading order analysis of the full nonlinear system restricted to the subspace associated with $ \lambda_{1,2} $ reveals a homoclinic which is, however, unlikely to persist for the full system. So, in general, one expects that resonances related to the infinitely many eigenvalues on the imaginary axis generate
oscillatory tails, hence, preventing the existence of truly localized breathers for the full PDE. The special structure of the sG-nonlinearity seems to remedy this spectral picture, allowing the homoclinic to persist. Nevertheless, other mechanisms to avoid such resonances are possible in KG equations with e.g. periodic coefficients \cite{bcls}.

\subsection{Spectral Properties of the Breather: From the Anti-Continuum
to Finite Coupling}
\jcmindex{\myidxeffect{A}!anti-continuum limit}
In a chain of $N$ oscillators described by e.g. Eq.~(\ref{cheqn2}) or (\ref{cheqn4}), discrete breathers in the AC limit consist of $p$ (excited) oscillators describing periodic orbits of frequency $\omega\neq\omega_0$ and $N-p$ oscillators at rest. Here, $\omega_0$ is the frequency of the oscillations at the linear limit around the equilibrium position $\tilde u$,
with $\tilde u=0$ in the sine-Gordon and $\tilde u=\pm1$ in the $\phi^4$ case. Breathers with $p>1$ are usually dubbed {\em multibreathers}; in the special case when $p=N$, they constitute an anharmonic phonon and are also known as {\em phonobreathers}. The MacKay--Aubry theorem \cite{macaub} establishes that if $n\omega\neq\omega_0$ with $n$ being an integer, a time-reversible breather or multibreather can be continued to a nonzero coupling (i.e. $\epsilon\neq0$). The time-reversible multibreather can be expressed as a Fourier series:
\jcmindex{\myidxeffect{D}!discrete breather}
\begin{equation}
    u_n(t)=\sum_{k=0}^\infty z_{k,n}\cos(k\omega t).
\label{cheqn28}
\end{equation}

In the case of the sine-Gordon equation, due to the spatial symmetry of the potential, all the even coefficients are zero and the resonance condition mentioned above restricts to only odd multiples of $\omega$.

Among all the localized
multibreathers supported by the lattice, there are only two that can be continued up to $\epsilon\rightarrow\infty$: (1) the single-site breather, i.e. the breather with $p=1$ (also known as the site-centered breather or Sievers--Takeno mode \cite{stak}), and (2) the two-site breather, i.e. multibreather with $p=2$, whose excited sites are adjacent and oscillate in phase, which is also dubbed as the bond-centered breather or Page mode \cite{page}.

Very recently, it was proven that non-time-reversible multibreathers cannot exist in Hamiltonian lattices if the number of neighbors of each oscillator is equal to two and the boundary conditions are not periodic \cite{kouk3}. This proof does not forbid, however, the existence of non-time-reversible multibreathers in two-dimensional lattices where discrete vortices or percolating clusters \cite{kouk2,ijbc,cretegny1,cretegny2} have been found in the past. In addition, non-time-reversible multibreathers have also been demonstrated in one-dimensional chains with periodic boundary conditions \cite{phason} or long-range interactions \cite{kouk4}.

In order to obtain periodic orbits of the lattice equations (\ref{cheqn2}) or (\ref{cheqn4}), a fixed-point
method (such as the Newton-Raphson method) can be implemented either in real or in Fourier space. In the real space case (see e.g. \cite{cretegny1}), orbits are found as roots of the map $(\{u_n(T)\},\{\dot u_n(T)\}) - (\{u_n(0)\},\{\dot u_n(0)\})$ with $T=2\pi/\omega$. The Jacobian of this map is  $\mathcal{M} - I$ where $\mathcal{M}$ is the
monodromy matrix (see e.g.~\eqref{cheqn32}) and $I$ is the identity.  Fourier space methods consist of introducing the Galerkin truncation to index $k_m$ of the Fourier series expansion (\ref{cheqn28}) into the lattice equations so that they can be cast as a set of algebraic equations of the form \cite{phason}:

\begin{equation}
    -k^2z_{k,n}^2=\epsilon\Delta_2 z_{k,n}-\mathcal{F}_{k,n},
\label{cheqn29}
\end{equation}
with $\mathcal{F}_{k,n}$ being the discrete cosine Fourier transform,

\begin{equation}
    \mathcal{F}_{k,n}=\frac{1}{N}\sum_{n=-k_m}^{k_m}V'\left(\sum_{p=-k_m}^{k_m}z_{p,n}\cos\left[\frac{2\pi p n}{N}\right]\right)
    \cos\left[\frac{2\pi k n}{N}\right],
\label{cheqn30}
\end{equation}
with $N=2k_m+1$ and $V'(u)=\sin(u)$ for the sine-Gordon model and $V'(u)=u^3-u$ for the $\phi^4$ equation\footnote{Notice in order to use the same notation as in classical papers as \cite{kruskalseg} and \cite{boyd}, the factor 2 that was introduced in (\ref{cheqn4}) has been removed. However, the results are equivalent by rescaling  $x\rightarrow\sqrt{2} x$ and
$t\rightarrow\sqrt{2}t$.}. Fourier space methods have the advantage of providing with an analytical expression for the Jacobian, whereas it must be numerically generated by means of numerical integrators (as Runge-Kutta) in real space methods. On the contrary, the dimension of the Jacobian in Fourier space $[(k_m+1)N,(k_m+1)N]$ is larger than in the real space $(N,N)$ which makes the inversion
intractable for large lattices.

Spectral stability of periodic orbits can be determined by means of Floquet analysis. This consists firstly of introducing a perturbation $\xi_n$ to a given solution $u_{n,0}$ of the lattice equations (\ref{cheqn2}), (\ref{cheqn4}). Then, the equation for the perturbation reads:

\begin{equation}
    \ddot \xi_n=\epsilon\Delta_2\xi_n-V''(u_{n,0}) \xi_n,
\label{cheqn31}
\end{equation}
with $V''(u_n)=\cos u_n$ and $V''(u_n)=3u_n^2-1$ for, respectively, the sine-Gordon and $\phi^4$ models. The stability properties are given by the spectrum of the Floquet operator $\mathcal{M}$ (whose matrix representation is called {\em monodromy}) defined as:
\jcmindex{\myidxeffect{M}!monodromy} \jcmindex{\myidxeffect{F}!Floquet operator}
\begin{equation}
    \left(\begin{array}{c} \{\xi_{n}(T)\} \\ \{\dot\xi_{n}(T)\} \\ \end{array}
    \right)=\mathcal{M}\left(\begin{array}{c} \{\xi_{n}(0)\} \\ \{\dot\xi_{n}(0)\} \\ \end{array}
    \right).
\label{cheqn32}
\end{equation}

The $2N$ monodromy eigenvalues $\Lambda=\exp(\mathrm{i}\theta)$ are dubbed as {\em Floquet multipliers} and $\theta$ are denoted as {\em Floquet exponents}. Due to the symplectic nature of the Floquet operator for our Hamiltonian systems
of interest, if $\Lambda$ is a multiplier, so is $\Lambda^{-1}$, and due to its real character if $\Lambda$ is a multiplier, so is $\Lambda^*$. Consequently, Floquet multipliers always come in complex quadruplets $(\Lambda,1/\Lambda,\Lambda^*,1/\Lambda^*)$ if $|\Lambda|\neq1$ and is not real, or in pairs $(\Lambda,1/\Lambda)$ if $\Lambda$ is real, or $(\Lambda,\Lambda^*)$ if $|\Lambda|=1$ and is not real. In addition, due to the time translation invariance of the system there is always a pair of multipliers at $+1$. Therefore, for a breather to be stable, all the multipliers must lie on the unit circle.

In the AC limit, Floquet multipliers lie in three bundles. The two conjugated ones, corresponding to the oscillators at rest, lie at $\theta=\pm2\pi\omega_0/\omega$, while the third one lies at $+1$ and consists of $p$ multiplier pairs corresponding to the excited oscillators. For $\epsilon\neq0$, the bundle corresponding to the oscillators at rest split and their multipliers move to form the phonon band, while the multipliers corresponding to the excited oscillators can move along, either in the unit circle (stability), or along the real axis (instability). However, as mentioned above, a pair of multipliers remains at $+1$ corresponding to the, so-called, phase and growth modes of the whole system. This implies that single-site breathers are stable for low coupling \cite{aubry}.

Whereas the stability at low coupling of single-site breathers is trivial, the situation for multibreathers
is less clear.

\begin{figure}
\begin{tabular}{cc}
\includegraphics[width=6cm]{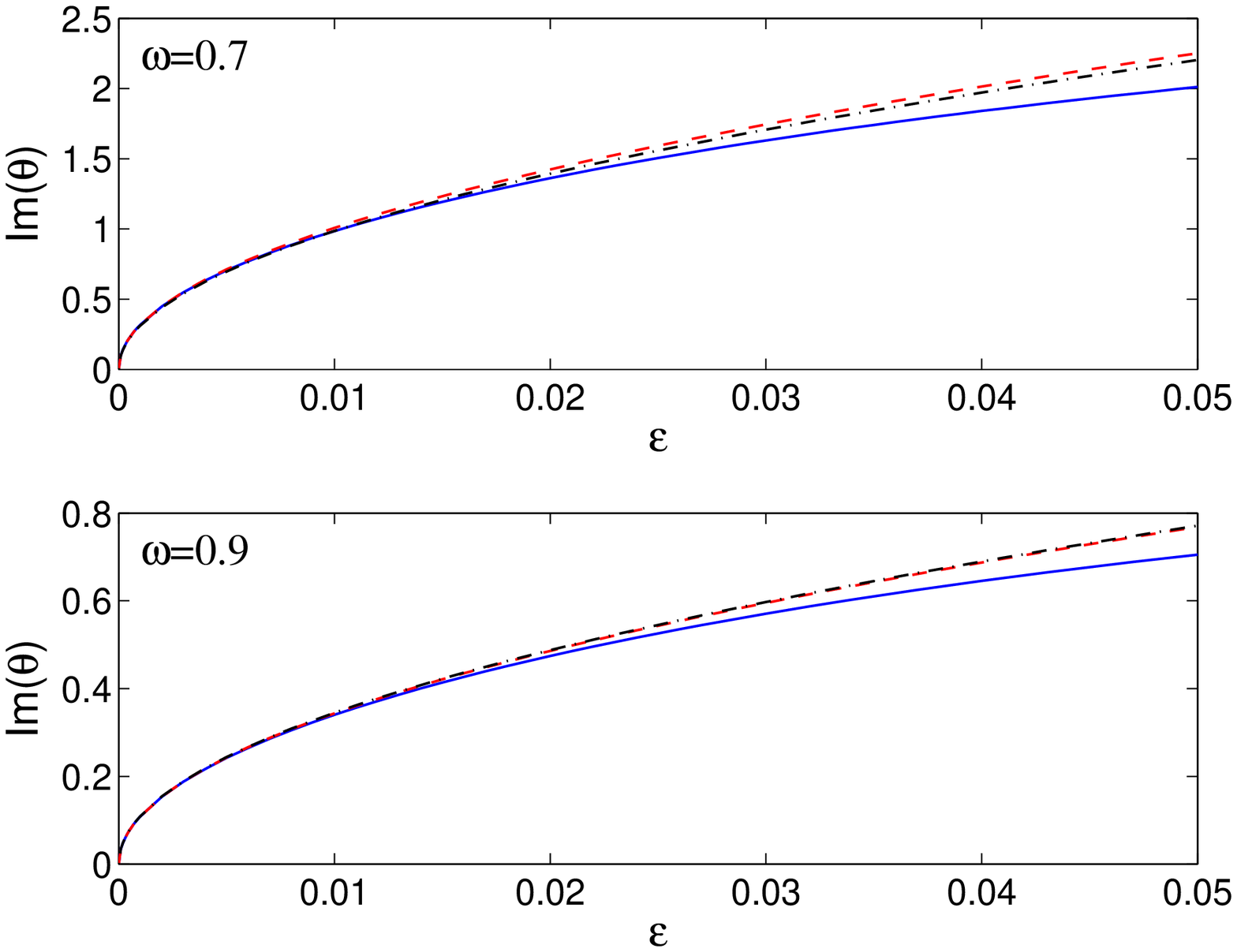}
\includegraphics[width=6cm]{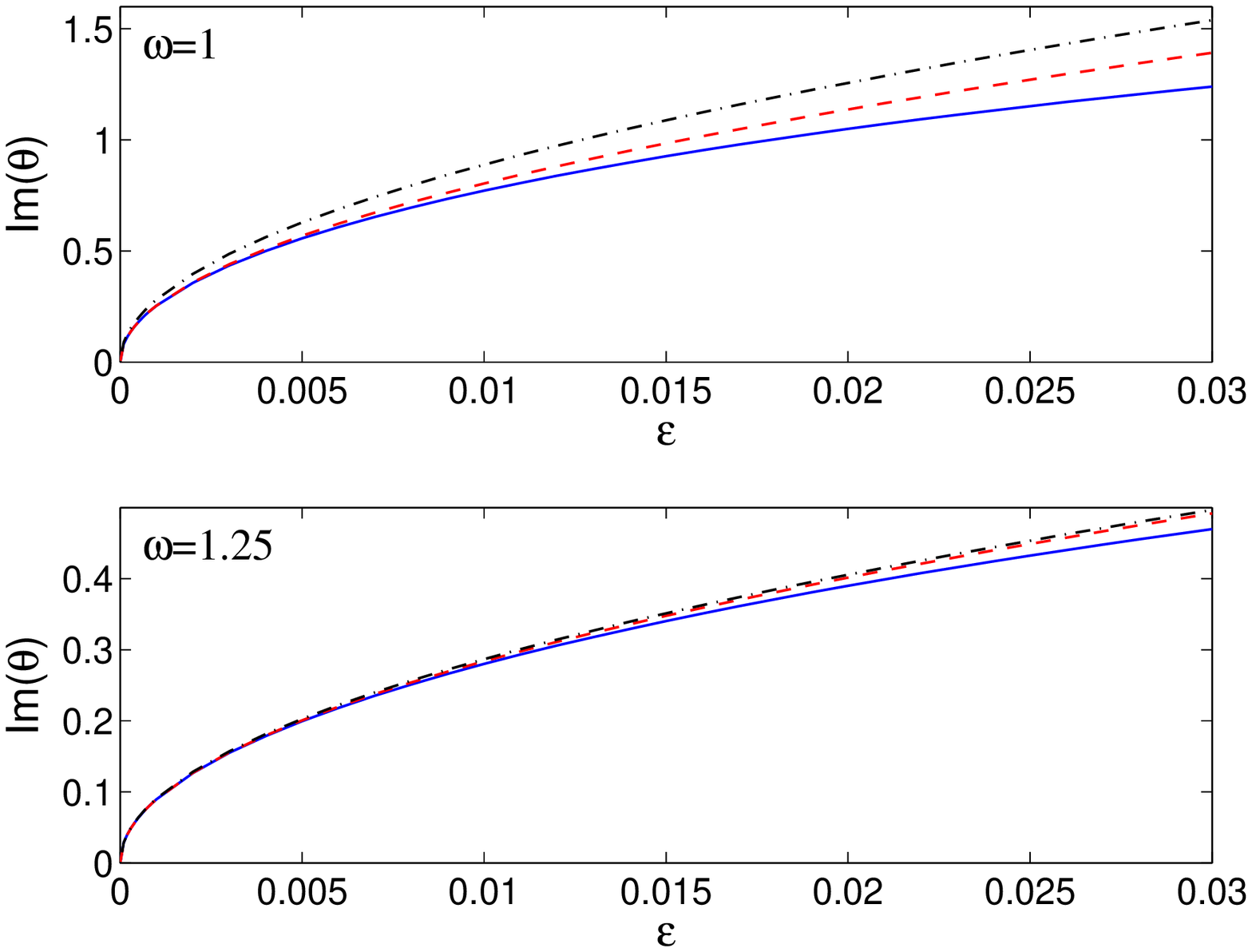}
\end{tabular}
\caption{Floquet exponent for the Page mode in the sine-Gordon (left) and $\phi^4$ (right) models. Full line corresponds to the numerically exact result, whereas the stability theorem predictions are given by the dashed line.
The explicit analytical predictions of the RWA approach
for the stability exponents are depicted by the dash-dotted line.}
\label{chfig6}
\end{figure}

The theorems developed in \cite{mst,kouk1} provide insight regarding this question.  One can derive an expression of the Floquet multipliers corresponding to the excited oscillators of multibreathers close to the AC limit. In the case of Page modes, the relevant Floquet exponent is given by:

\begin{equation}
    \theta=\pm \frac{2\pi}{\omega}\sqrt{2\epsilon\frac{J}{\omega}\frac{\partial \omega}{\partial J}},
    \label{cheqn33}
\end{equation}
with
\begin{equation}
    J=\frac{1}{2\pi}\int_0^T[\dot u(t)]^2\mathrm{d}t=\frac{\omega}{2}\sum_{k\geq1} k^2z_k^2,
    \label{cheqn34}
\end{equation}
being the action of an isolated oscillator. Applying a rotating wave approximation (RWA) $u(t)\approx z_0+z_1\cos(\omega t)$
we obtain the following Fourier coefficients for the sG model \cite{abram}:

\begin{equation}
    z_0=0,\qquad z_1=12-4\sqrt{3[3-4(1-\omega^2)]},
    \label{cheqn35}
\end{equation}
whereas the following coefficients are found for the $\phi^4$ model:
\begin{equation}
    z_0=1-\frac{3}{5}(2-\omega^2), \qquad z_1=\frac{2}{5}(2-\omega^2).
    \label{cheqn36}
\end{equation}
Using the values from (\ref{cheqn33}), the relevant Floquet exponent at low coupling for the sG model can now be approximated by:

\begin{equation}
    \theta=\pm \frac{2\pi}{\omega}i\sqrt{2\epsilon\frac{\omega^2\sqrt{3(4\omega^2-1)}-(2\omega^2-1)(4\omega^2-1)}{16\omega^4-7\omega^2+1}},
    \label{cheqn37}
\end{equation}
whereas for the $\phi^4$ equation we have
\begin{equation}
    \theta=\pm \frac{2\pi}{\omega}i\sqrt{2\epsilon\frac{\omega^2-2}{2-3\omega^2}}.
    \label{cheqn38}
\end{equation}
Figure~\ref{chfig6} compares the Floquet exponents of the Page mode found by integrating the perturbation equation (\ref{cheqn32}) with those of the analytical approximation (\ref{cheqn33}) and the subsequent, fully analytical RWA (\ref{cheqn37})-(\ref{cheqn38}) for low coupling. Good agreement is found in the
case of small values of the coupling constant $\epsilon$.

\subsection{Continuation in the Coupling: Going towards the Continuum}

\begin{figure}
\begin{tabular}{cc}
\includegraphics[width=6cm]{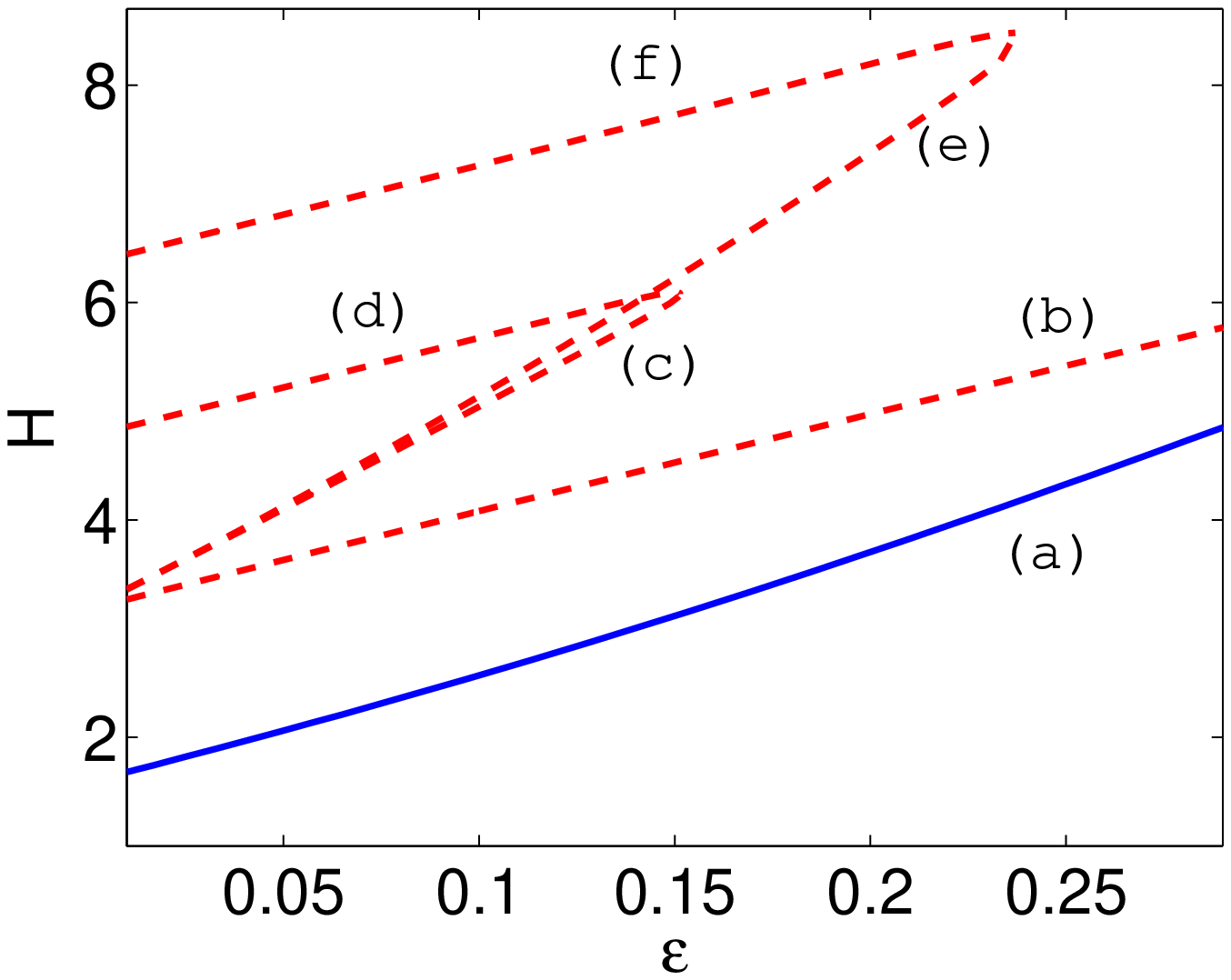}
\includegraphics[width=6cm]{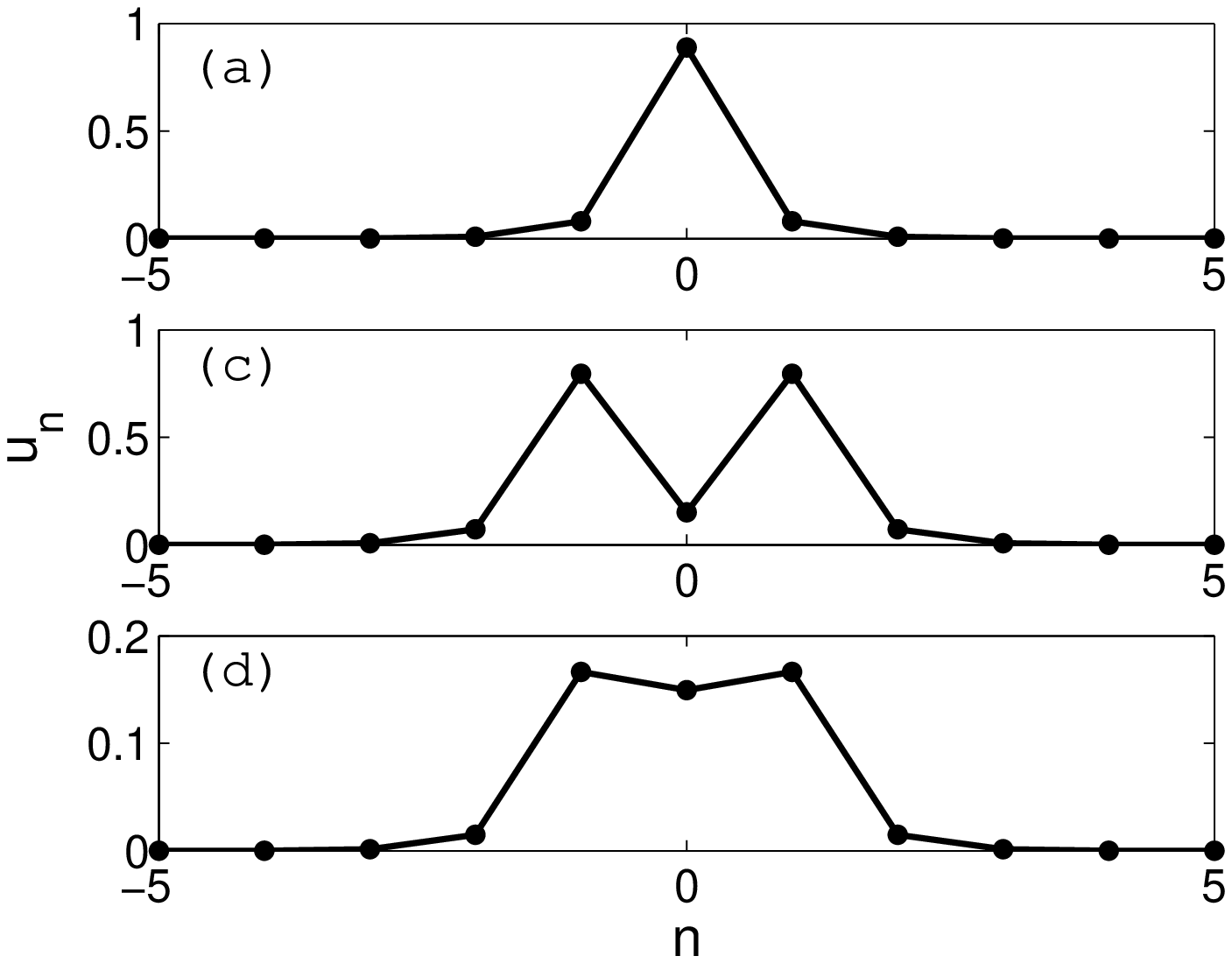}
\end{tabular}
\caption{Left: Hamiltonian energy (H) of breathers of the discrete sG
equation as the coupling parameter $\epsilon$ is increased for $\omega = 0.7$. The labels correspond to breathers which have the following
excited nodes in the AC limit: (a) $ (\ldots,\bullet,\bullet,\bullet,\uparrow,\bullet,\bullet,\bullet,\ldots) $ (b) $(\ldots,\bullet,\bullet,\bullet,\uparrow,\uparrow,\bullet,\bullet,\ldots) $ (c) $(\ldots,\bullet,\bullet,\uparrow,\bullet,\uparrow,\bullet,\bullet,\ldots) $ (d) ($ \ldots,\bullet,\bullet,\uparrow,\uparrow,\uparrow,\bullet,\bullet,\ldots$) (e) $ (\ldots,\bullet,\bullet,\uparrow,\bullet,\bullet,\uparrow,\bullet,\bullet, \ldots) $ and (f) $ (\ldots,\bullet,\bullet,\uparrow,\uparrow,\uparrow,\uparrow,\bullet,\bullet,\ldots $). Here, arrows correspond to the excited sites, which oscillates with frequency $\omega$, whereas the oscillators at rest are represented by dots. Only the site-centered (a) and bond-centered (b) solutions survive
for $\epsilon \rightarrow \infty$, both of which approach the exact breather solution of the continuous sG equation, see Eq.~\eqref{EQ:breather_formula}. Right: Examples of discrete sG breathers for $\epsilon=0.065$. The multibreathers in the  two bottom panels collide in a fold bifurcation
at $\epsilon \approx 0.152$. The labels correspond to those in the left panel. }
\label{chfig_bif}
\end{figure}

When the coupling constant is increased to finite values of $\epsilon>0$, stable multibreathers that are close to the AC limit destabilize generally by means of Hamiltonian Hopf bifurcations (also referred to as Krein crunches); in addition, every multibreather ceases to exist for a finite value of the coupling constant by means of Hopf or fold bifurcations. As discussed above, the
two exceptions to these facts are given by the Sievers-Takeno (site-centered breathers) and Page modes (bond-centered breathers) which can be continued up to the continuum limit ($\epsilon\rightarrow\infty)$. During this continuation, however, both modes experience important changes both in their shape and in their
spectral properties, as we will see below.

The site-centered and bond-centered modes can be continued until resonance with the linear phonon band occurs. For periodic boundary conditions and assuming
that the lattice has an {\em even} number of particles $N$, the phonon band is given by:
\begin{equation}
    \omega^2_\mathrm{ph}(\epsilon)=\omega_0^2+4\epsilon\sin^2\frac{q}{2},\qquad\mathrm{with }\qquad q=\left(1-\frac{2n}{N}\right)\pi,\ n=0,1,2\ldots N/2
    \label{cheqn39}
\end{equation}
where the modes are sorted by frequency and those from $n=1$ to $n=N/2-1$ are doubly degenerate; $\omega_0^2=1$ ($\omega_0^2=2$) in the sG ($\phi^4$) model.
Therefore, the first resonance, upon variation of
$\epsilon$ for fixed $\omega$, must occur when $n=0$, i.e. with the $q=\pi$ linear mode.
In the case of $\phi^4$ potential,  resonances occur when the second harmonic of the breather frequency coincides with a phonon frequency ($m\omega=\omega_\mathrm{ph}$, $m=2$), i.e. the first resonance occurs at $\epsilon_\pi=\omega^2-1/2$. However, in the sine-Gordon potential, the Fourier coefficients corresponding to even harmonics are zero, so that the resonance is caused by the third harmonic ($m\omega=\omega_\mathrm{ph}$, $m=3$) and the first resonance is observed at $\epsilon_\pi=(9\omega^2-1)/4$.

Thus, if a smooth continuation is performed, the site- (or bond-) centered solution continuously
transforms into a phonobreather (i.e., resonates with the linear
modes) once the coupling reaches the value $\epsilon_\pi$.
The central site(s) will oscillate with frequency $\omega$ whereas the
tail will
oscillate with frequency $m\omega$.  Nevertheless, if the step in the continuation method is large enough, the system is able to ``jump'' to a localized solution whose frequency lies in the gaps of the phonon band. Those solutions are the result of a hybridization of a breather with a phonon and are dubbed as {\em phantom breathers} \cite{phantom} \footnote{In the nonlinear Schr\"odinger equation framework those solutions have been called {\em hybrid lattice solitons} \cite{wsl}.}. Phantom breathers are in fact the discrete version of the \jcmindex{\myidxeffect{N}!nanopteron} nanopterons observed in the continuum $\phi^4$ equation \cite{boyd}. However, unlike the continuum models, they typically arise for on-site potentials (including e.g. the discrete
sine-Gordon) as long as the coupling is finite and higher than $\epsilon_\pi$.
Similarly to nanopterons and the phonobreathers found by smooth continuation, phantom breathers are composed of a localized breather oscillating with frequency $\omega$ plus a low-amplitude background of frequency $m\omega$.

\begin{figure}
\begin{tabular}{cc}
\includegraphics[width=5.5cm]{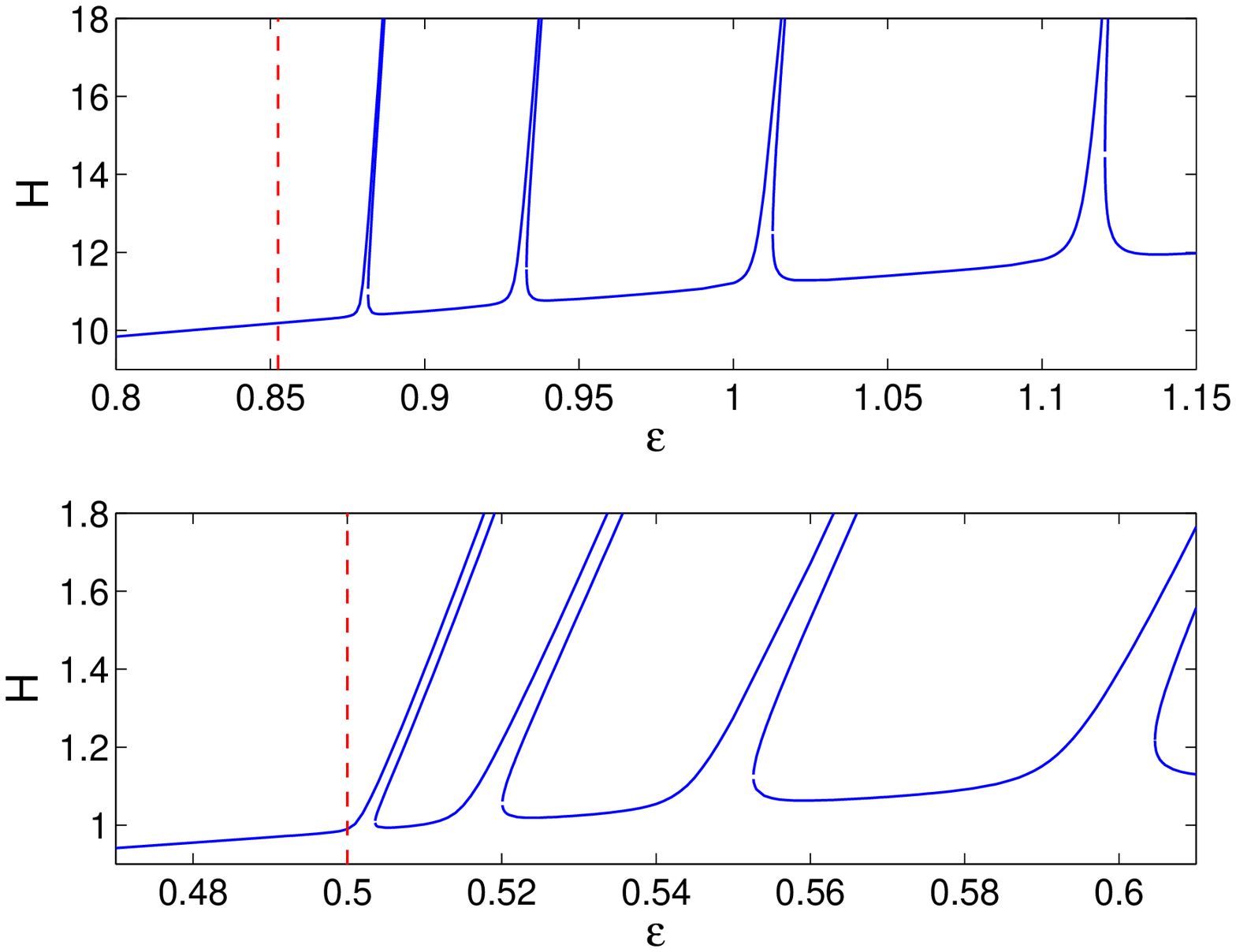}
\includegraphics[width=5.5cm]{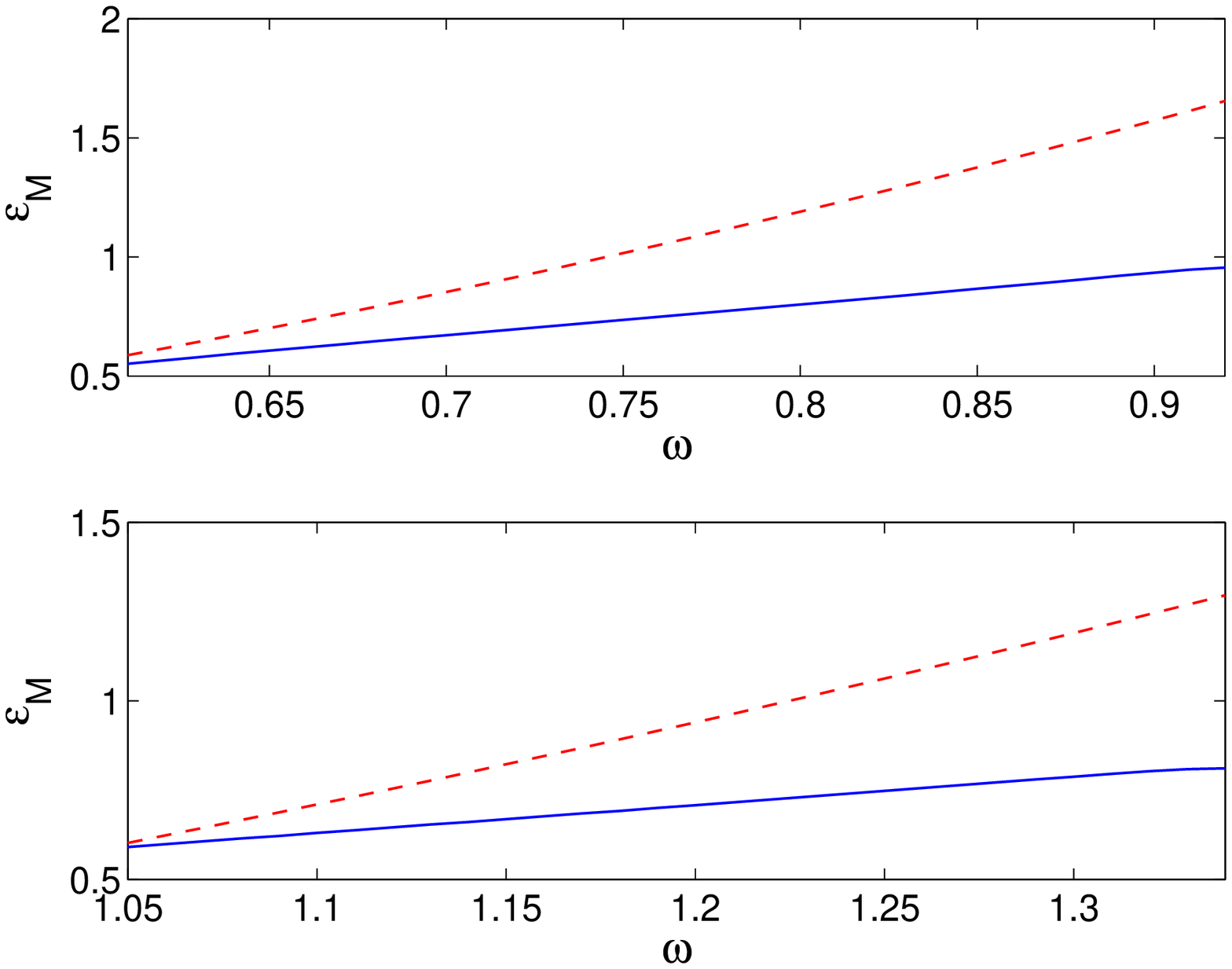}
\end{tabular}
\caption{(Left panel) Hamiltonian energy versus coupling for breathers with frequency $\omega=0.7$ in the sine-Gordon model (top) and $\omega=1$ for the $\phi^4$ model (bottom). The vertical dashed line indicates the value $\epsilon_\pi$.
(Right panel) Exchange of stability bifurcation loci for breathers in the sG (top) and $\phi^4$ potential (bottom); the dashed line indicates the value of $\epsilon_\pi(\omega)$.}
\label{chfig7}
\end{figure}

Let us mention that the bifurcation diagram for phantom breathers (see \cite{phantom} for an extensive study of such solutions) resembles the Wannier--Stark ladders observed in a model of non-homogenous waveguide arrays \cite{wsl} [see Fig. \ref{chfig7} (left)]; this implies the existence of three nonlinear modes for the same value of the coupling constant.
Of course, the width of gaps in the phonon band decreases with the number of system particles $N$, and the gaps eventually vanish for $N\rightarrow\infty$.
In other words, the hybridization of the breather
with the background phonons is a finite-size artifact that disappears
in the infinite lattice limit.

Similar to the kink situation described in Sec.~\ref{sec:kinkmerge}, there are important details regarding the spectral properties of discrete breathers in the $\phi^4$ and sine-Gordon lattices. In both models, the site- and bond-centered modes interchange their stability at a finite value of the coupling constant, which we denote as $\epsilon_M$. \jcmindex{\myidxeffect{S}!stability exchange}
This exchange-of-stability bifurcation (which is mediated by an intermediate breather) is closely related to the existence of {\em moving breathers} \cite{chen,cretegny3}. This bifurcation is caused by a spatially anti-symmetric internal mode called {\em pinning mode}. For the site-centered breather mode, this internal mode detaches from the phonon band and becomes localized when $\epsilon$ is increased, colliding at $\theta=0$ when the exchange of stability takes place. In addition, moving breathers not only exist close to the parameter values at which stability exchange takes place but also when the Floquet multiplier corresponding to the pinning mode is close to
1. In this case,  the movement is possible because the height of the Peierls--Nabarro barrier is related to the distance of the unstable multiplier to 1. Figure \ref{chfig7} (right) shows the dependence of $\epsilon_M$ with respect to $\omega$. In general, when $\epsilon$ becomes larger than $\epsilon_M$ the growth rate (distance of the unstable Floquet multiplier to 1) of the Sievers-Takeno mode grows up to a maximum value; beyond that point, there are many windows where stability exchanges again. Those windows correspond to the jumps into the phonon band and are difficult to control in the numerics. Nevertheless, as we approach $\epsilon \rightarrow \infty$, i.e., at the continuum limit, the relevant translational eigenvalue returns to the point $(1,0)$ of the unit circle, signaling
the restoration of the translational invariance in that limit.

\section{A Recent Variant: sine-Gordon Equations with $\mathcal{PT}$ Symmetry}
\jcmindex{\myidxeffect{P}!PT-symmetry}
One of the interesting recent extensions that have emerged in the
context of Klein-Gordon field theories is that of $\mathcal{PT}$-symmetric
variants of these models. The analysis of Hamiltonians respecting
the concurrent application of parity ($x \rightarrow -x$) and
time-reversal ($t \rightarrow -t$) operations was initially
proposed in a series of publications by Bender and
collaborators~\cite{bend1,bend2,bend3}.
While originally
intended for Schr{\"o}dinger Hamiltonians
as a modification/extension of quantum mechanics,
its purview has gradually become substantially wider.
This stemmed to a considerable degree from the realization
that other areas such as optics might be ideally suited not
only for the theoretical study of this theme~\cite{ziad,Ramezani,Muga},
but also for its experimental realization~\cite{salamo,dncnat,whisper}.
Recently, both at the mechanical
level~\cite{bend_mech} and at the electrical one~\cite{R21,tsampas2}
realizations of $\mathcal{P T}$ symmetry and its breaking
have arisen,
while a Klein-Gordon setting has also been explored theoretically
for so-called $\mathcal{P T}$ symmetric nonlinear metamaterials and
the formation of gain-driven discrete breathers therein~\cite{tsironis}.

These efforts have, in turn, prompted a number of works both
at the discrete level~\cite{ptkg,demirkaya}, as well as at
the continuum one~\cite{demirkaya,demirkaya2} to identify prototypical
settings of application of $\mathcal{PT}$-symmetry in nonlinear
KG field theories. The model used in~\cite{demirkaya,demirkaya2}
is of the form:
\begin{eqnarray}
 \label{cheqn22}
u_{tt}-u_{xx} +\epsilon \gamma(x) u_t   + f(u) = 0.
\end{eqnarray}
Here, in order to preserve the $\mathcal{P T}$ symmetry,
$\gamma(x)$, should be an antisymmetric function satisfying
$\gamma(-x)=-\gamma(x)$. Physically, this implies that while
this is an ``open'' system with gain and loss, the gain balances
the loss, in preserving the symmetry.

Interestingly, for this class of models, the stationary states
$u_0$, such as the kinks, are {\it unaffected} by the presence
of the $\mathcal{P T}$ symmetry (since the relevant term involves
$u_t$). On the other hand, the linearization problem becomes
\begin{equation}
 \label{cheqn23}
 \lambda^2 v +\epsilon \lambda \gamma v + H v =0.
 \end{equation}
where $H$ is the Hamiltonian linearization operator of e.g.
Eqs.~(\ref{cheqn9}) or (\ref{cheqn12}). Equivalently, the
spectral problem can be written as:
\begin{equation}
\label{cheqn24}
\lambda \left( \begin{array}{c} v_1 \\ v_2 \end{array}\right)=
\left( \begin{array}{cc} 0 & 1  \\
- H & -\epsilon \gamma(x)   \end{array}\right)\left( \begin{array}{c} v_1 \\ v_2 \end{array}\right).
\end{equation}

Now, the key question becomes what will be the fate of the eigenstates
associated with the linearization around the continuum kink, in the
presence of the perturbation imposed by the $\mathcal{PT}$-symmetric
gain/loss term. What has been argued in~\cite{demirkaya,demirkaya2}
is that in both the continuum and the infinite lattice discrete problem,
the continuous
spectrum will not be altered. Hence, the question is what becomes
of the point spectrum in the presence of this perturbation.

\begin{figure}
\begin{tabular}{cc}
\includegraphics[width=5.5cm]{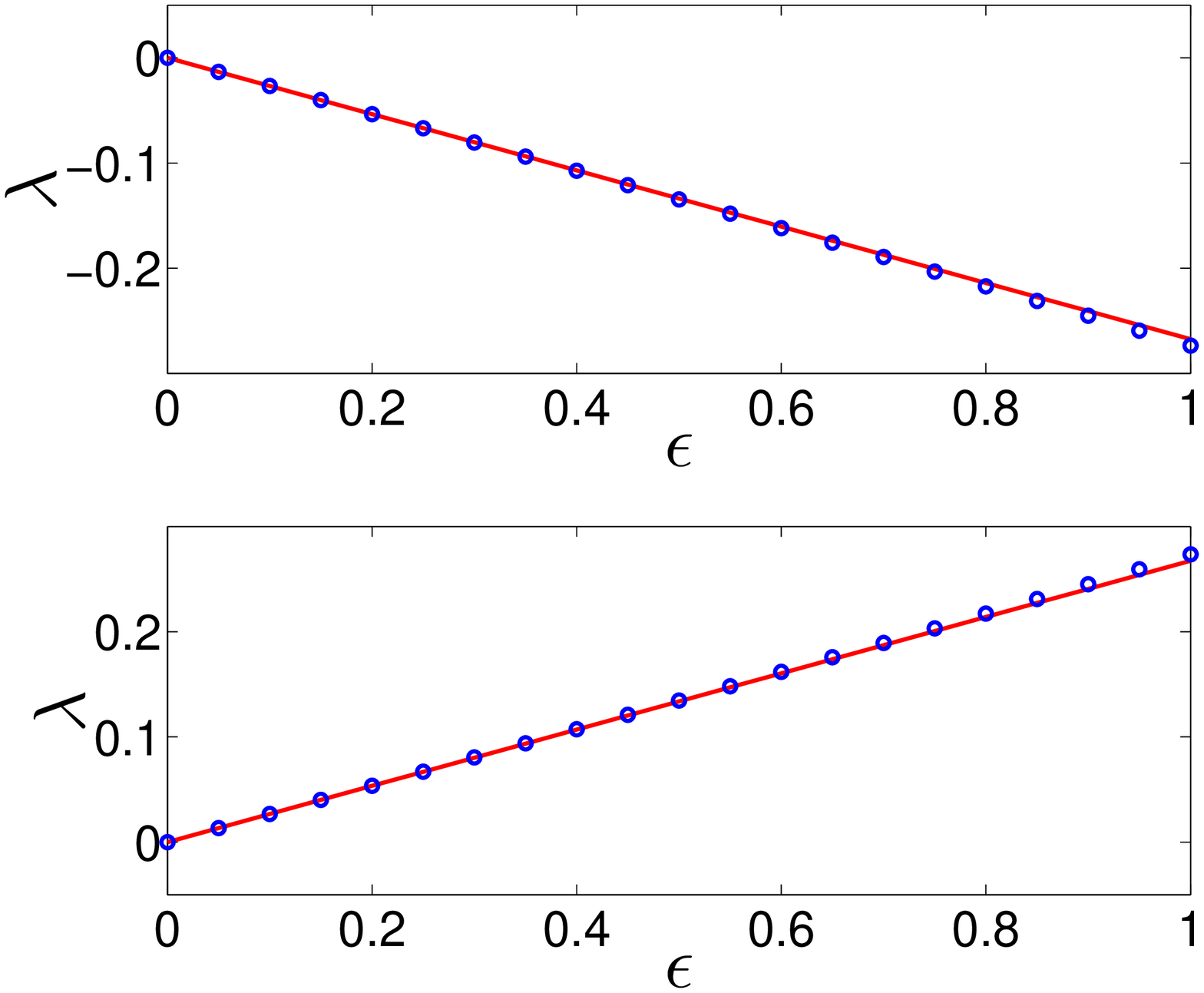}
\includegraphics[width=5.5cm]{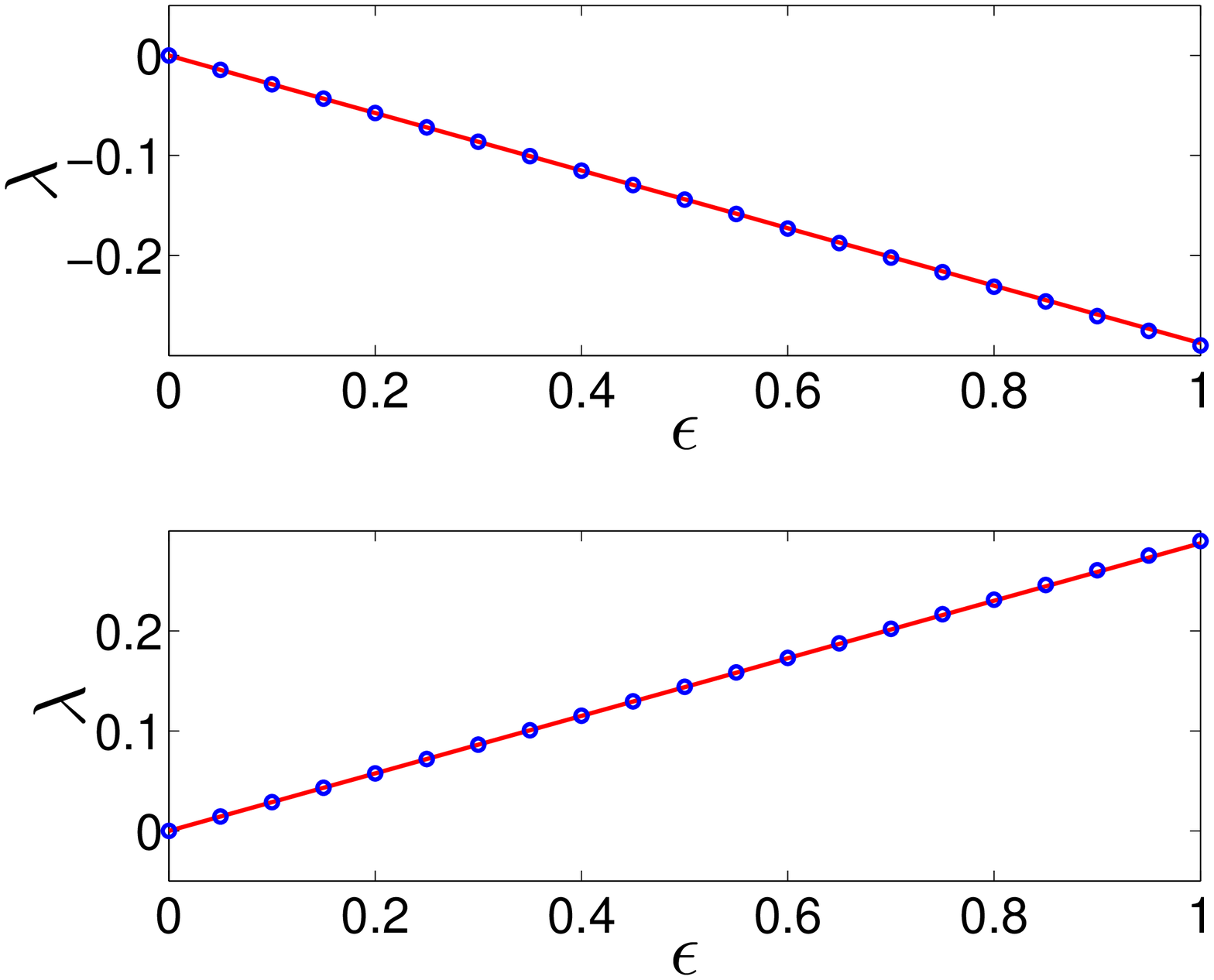}
\end{tabular}
\caption{A typical example of the comparison of the motion
of the translational eigenvalue under a $\mathcal{PT}$-symmetric
perturbation between numerics (points) and analytics of Eq.~(\ref{cheqn25}).
The left is for sG and the right for $\phi^4$, the top for a kink
centered at the lossy side, while the bottom for a kink centered
at the gain side.}
\label{chfig4}
\end{figure}

To determine what happens to the mode associated with translation,
we project Eq.~(\ref{cheqn23}) to the eigenfunction of vanishing eigenvalue
(in the unperturbed limit $v = u_{0,x}$), and observe that
the last term in Eq.~(\ref{cheqn23}) disappears. Assuming a perturbative
expansion of $v$ in $\epsilon$, which to leading order
(when $\epsilon=0$) has $v= u_x$, leads immediately to the
leading order approximation
\begin{eqnarray}
\lambda=0 \quad {\rm or} \quad \lambda=- \epsilon
\frac{\int \gamma(x) u_{0,x}^2 dx}{\int u_{0,x}^2 dx}.
\label{cheqn25}
\end{eqnarray}
This explicit expression should yield a fairly accurate prediction
of the fate of the former double $0$ eigenvalue pair.
Interpreting this result, we find that one of the two eigenvalues
of the pair should remain at $0$ (a result that persists to all
orders in perturbation theory due to the remaining invariance of the
location of the solution under translations). On the other hand,
the fate of the second eigenvalue will depend on the
relative position of the kink centered at $x_0$ with respect to the origin.
In particular, for both the sine-Gordon and the $\phi^4$, we
find that if the kink is centered at the lossy side (say, $x_0<0$),
then the numerator of the expression in Eq.~(\ref{cheqn25})
is negative, leading the second eigenvalue to the left hand
plane. Reversing the sign of $x_0$ leads effectively to a reversal
of the time flow (due to the nature of the system) and results
into the corresponding eigenvalue moving by an equal amount to the
right half of the spectral plane i.e., yielding instability.
On the other hand, for the special case in which $x_0=0$,
the parity of the relevant integral makes it vanish and
the eigenvalue does not shift. A typical example showcasing
the quality of the agreement of the theoretical prediction
with full numerical results is shown in Fig.~\ref{chfig4}.

However, in the case of the $\phi^4$ model, it is not sufficient
to consider only the fate of the eigenvalue potentially bifurcating
from the origin of the spectral plane. Recall that there is
an eigenvalue at $\lambda^2=-3$.
One can consider a perturbative expansion of such an eigenvalue
in the problem, according to $\lambda= \lambda_0 + \epsilon \lambda_1
+ \dots$, while $v=v_0 + \epsilon v_1 + \dots$. Applying a solvability condition at the leading order of the perturbation
theory then immediately yields the eigenvalue leading order correction
as:
\begin{eqnarray}
\lambda_1 = -\frac{1}{2} \frac{\int \gamma(x) v_0^2 dx}{\int v_0^2 dx}.
\label{cheqn26}
\end{eqnarray}
The interpretation of this result is that this
isolated point
spectrum eigenvalue (pair) will also move to the left half plane,
for the case of the kink centered at the lossy side, while it
will move (as a complex pair) to the right half plane when the kink is
centered at the gain side. I.e., overall stability of the kink's
point spectrum is ensured when centered at the lossy region and
instability when centered at the gain region. When the kink is at
the interface, it should already be clear from the above considerations
that all real contributions to the eigenvalue will vanish, allowing
only imaginary ones that could shift the point spectrum along the
imaginary axis, as proved earlier in the text.
Again, very good agreement between this prediction and the full
numerical result was found in~\cite{demirkaya2}.

\begin{figure}
\begin{tabular}{cc}
\includegraphics[width=6cm]{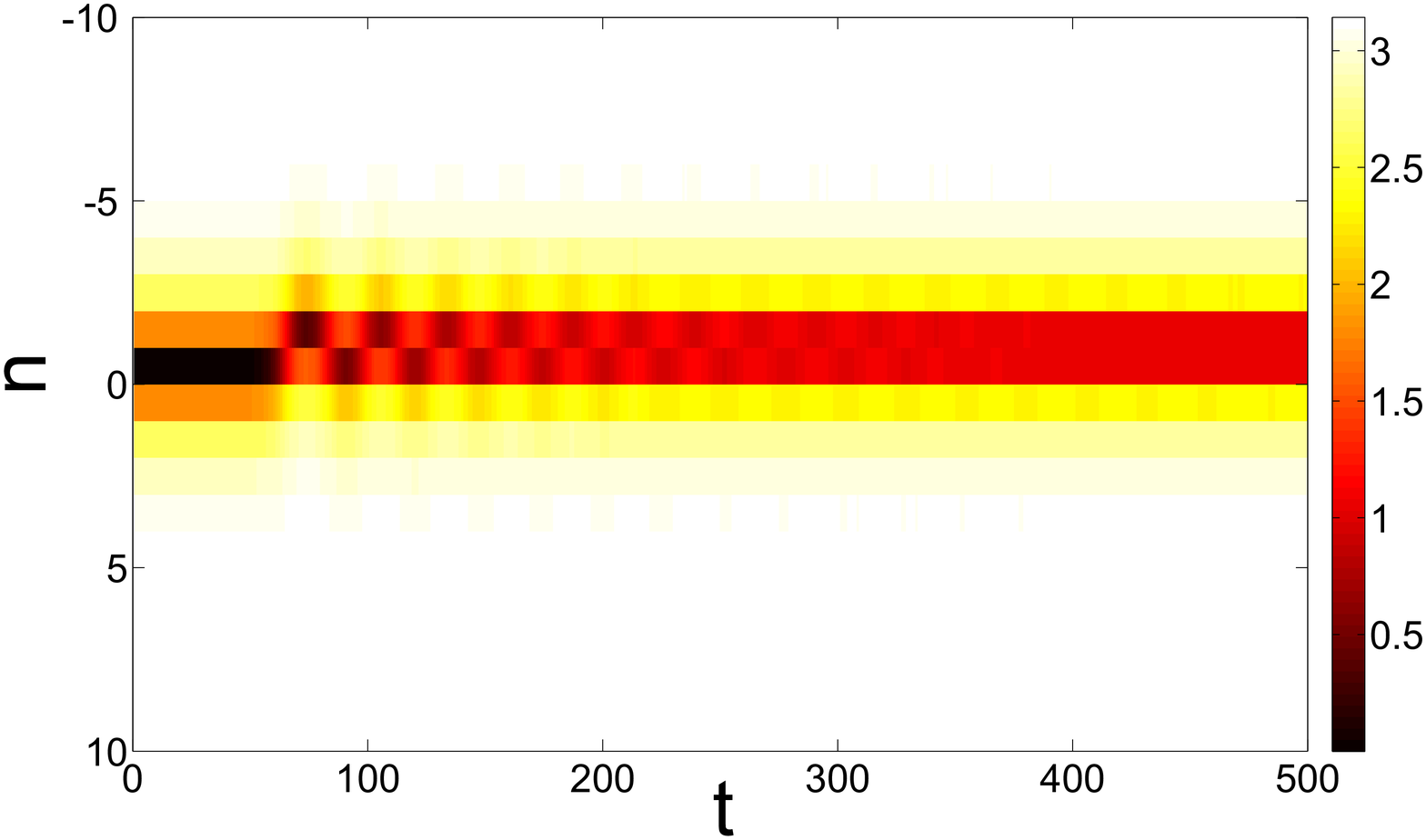}
\includegraphics[width=6cm]{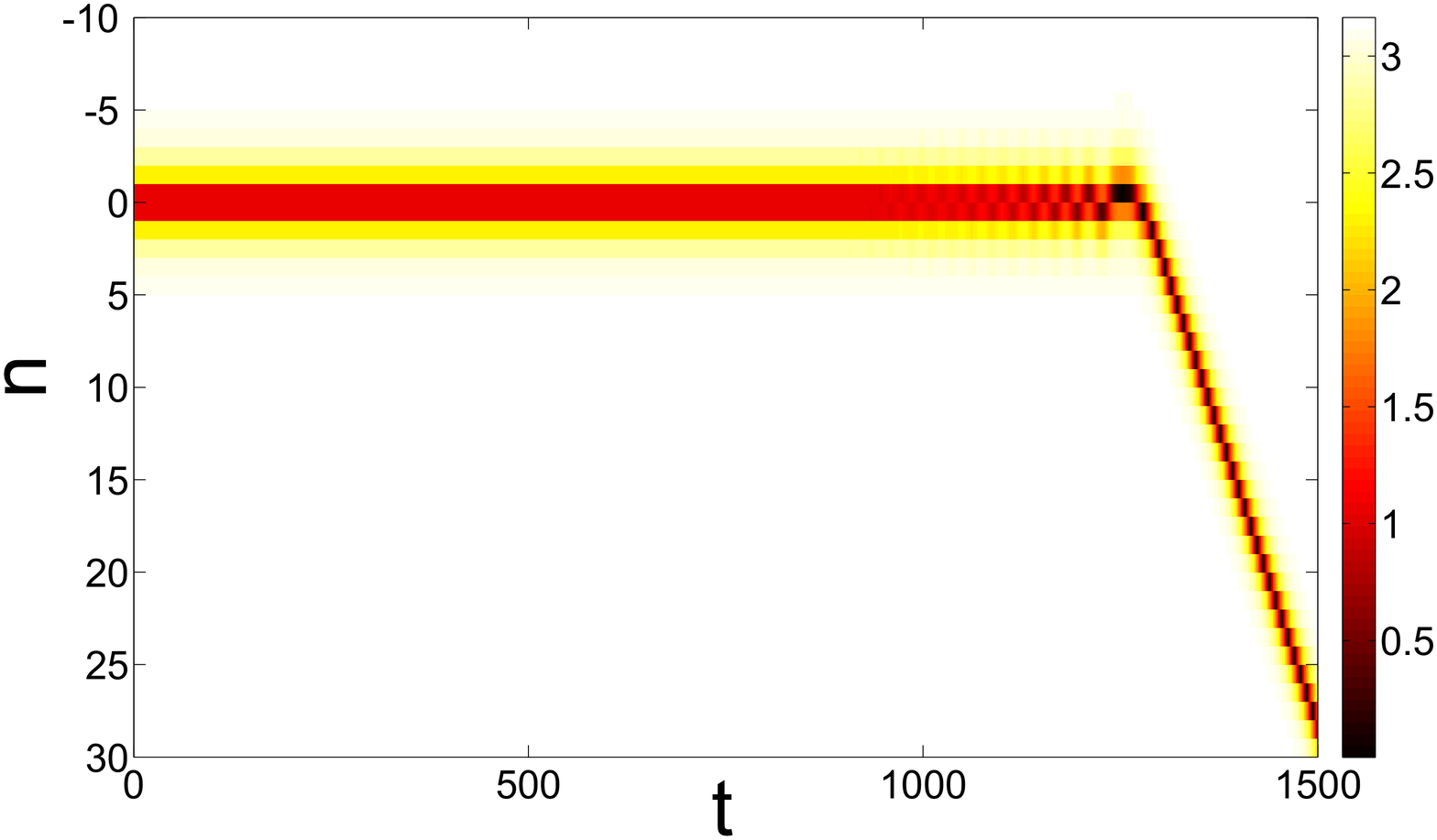}
\end{tabular}
\caption{Two canonical examples of the evolution of an unstable
kink on the gain side of a discrete $\mathcal{PT}$-symmetrically
perturbed sG model. The left panel shows the kink as settling into
an intersite-centered structure on the lossy side, while the right
forces it to propagate on the gain side, but without bearing a stable
fixed point for the kink state. Color bar represents $|u_n-\pi|$.}
\label{chfig5}
\end{figure}

In the discrete case, the same methods as explored above yield
equally accurate predictions. Arguably, the only significant
modification arising here is in the case of the intersite centered
stable kink. For the onsite one, the perturbation shifts the
eigenvalue along the real axis (weakly, for small perturbation
strength), not changing the fate of the kink's (in)stability.
However, in the marginal case of the intersite centered kink,
once again, the lossy side leads to stability kicking the
former translational (now imaginary in the discrete problem)
eigenvalues to the left half-plane. Placing the kink on the gain
side has the opposite effect with eigenvalues bifurcating on the
right half plane (as a complex pair) and hence leading to instability.
It is for that reason that the fate of the unstable discrete kink
of the dynamical example of Fig.~\ref{chfig5} leads to a static intersite
centered structure on the lossy (left) side for the left panel
of the figure. If, however, the perturbation pushes (intentionally)
the kink toward the right hand (gain) side, it cannot find a stable equilibrium
point and instead it continues to propagate as is shown in the
right panel of the figure.

In the case of breathers, we can add that also based on the discussion
given earlier for different variants of the sG equation and KG
equations more generally, breathers are unlikely to exist under
PT-symmetric perturbations. However, we do find that in the case
example potentials that we considered, stationary breather solutions
can only be found to exist at the delicate interface between
gain and loss. Breathers seeded on the lossy side are found to
become mobile. Perhaps more interestingly, such structures on the
gain side appear to gain energy and transform themselves into
a kink-antikink pair.

\section*{Acknowledgments}

PGK gratefully acknowledges support from grants NSF-CMMI-1000337, NSF-DMS-1312856, US-AFOSR FA-9550-12-1-0332,
BSF-2010239 and from the ERC through an IRSES grant. The work of MCB was partially supported by the Deutsche Forschungsgemeinschaft (DFG) under the grant CH 957/1-1. We acknowledge Aslihan Demirkaya for her technical assistance.

\end{document}